\documentclass[12pt]{article}
\usepackage{epsfig}
\usepackage{graphicx}
\usepackage{amsmath, amsfonts}
\usepackage{amssymb}
\usepackage{latexsym}
\usepackage{color}
\usepackage{mathrsfs}
\usepackage{natbib}
\usepackage{hyperref}
\usepackage{appendix}
\usepackage{dsfont}
\usepackage[left=3cm,top=4cm,right=3cm,bottom=3cm]{geometry}
\RequirePackage[caption=false]{subfig}
\usepackage{lscape}
\usepackage{dsfont}
\usepackage{rotating}

\usepackage{mathabx}

\graphicspath{{./figs/}}

\newcommand{\cb}{\mathbf{c}}
\newcommand{\de}{\mathrm{d}}
\newcommand{\e}{\mathds{E}}
\newcommand{\R}{\mathds{R}}
\newcommand{\X}{\Breve{X}}
\newcommand{\M}{\mathds{M}}
\newcommand{\si}{\mathds{S}}

\newcommand{\oplust}{\oplus_{\mathds{T}}}
\newcommand{\odott}{\odot_{\mathds{T}}}
\newcommand{\rt}{\rangle_{\mathds{T}}}
\newcommand{\tf}{\mathfrak{t}}
\newcommand{\var}{\mathds{V}\text{ar}}

\DeclareMathOperator{\cls}{\mathrm{cls}}
\DeclareMathOperator{\alr}{\mathrm{alr}}

\DeclareMathOperator{\clr}{\mathrm{clr}}
\DeclareMathOperator{\ilr}{\mathrm{ilr}}

\DeclareMathOperator{\clrinv}{\mathrm{clr}^{-1}}

\DeclareMathOperator{\ailr}{i\alpha t}
\DeclareMathOperator{\aclr}{c\alpha t}
\DeclareMathOperator{\NT}{\overline{\mathbf{T}}}

\DeclareMathOperator{\cen}{cen}

\DeclareMathOperator{\mvar}{m\mathds{V}ar}

\newcommand{\h}{\mathbf{H}_{\mathrm{D}}}

\begin{document}
\title{\LARGE \bf On spatial point processes with composition-valued marks}
\date{\date{}}
\maketitle
\begin{center}
{{\bf Matthias Eckardt$^{a}$}, {\bf Mari Myllym{\"a}ki}$^{b}$} and {\bf Sonja Greven$^{a}$}\\
\noindent $^{\text{a}}$ Chair of Statistics, Humboldt-Universit\"{a}t zu Berlin, Berlin, Germany\\
\noindent $^{\text{b}}$ Natural Resources Institute Finland (Luke), Helsinki, Finland 
\end{center}

\begin{abstract}

Methods for marked spatial point processes with scalar marks have seen extensive development in recent years. 
While the impressive progress in data collection and storage capacities has yielded an immense increase in  spatial point process data with highly challenging non-scalar marks, methods for their analysis are not equally well developed. 
In particular, there are no methods for composition-valued marks, i.e.\ vector-valued marks with a sum-to-constant constrain (typically 1 or 100). Prompted by the need for a suitable methodological framework, we extend  existing  methods to spatial point processes with  composition-valued marks and adapt common mark characteristics to this context. The proposed methods are applied to analyse spatial correlations in data on tree crown-to-base and business sector compositions.    
\end{abstract}
\section{Introduction}

In recent years, there has been substantial interest in the analysis of marked spatial point processes. This type of data involves random spatial positions of $n$ events with marks providing specific additional details about each event. Although spatial point process theory is mature, and various summaries for marked points have been created \citep[see][]{StoyanStoyan1994, Illian2008, Chiu2013}, there is still a gap in addressing non-scalar marks. Particularly, spatial point processes with compositional marks—where marks are made up of $D$ components that together sum to a constant—are not fully examined. Such instances include pest infestation ratios in wood samples, proportions of business sectors locally, and parts of tree biomass. Here, marks present relative information with interdependent $D$ components; an increase in one results in a decrease in others, due to the fixed sum. The constrained nature of this data demands analysis techniques beyond traditional Euclidean methods. Individual component analysis within the current framework disregards these constraints, risking biased or spurious outcomes \citep{https://doi.org/10.1029/JZ065i012p04185, pawlowsky1984spurious}. Thus, a methodological framework for analyzing such composition-valued marks remains crucial in modern applications. This paper seeks to address this by introducing a new class of spatial point processes with composition-valued marks and expanding the toolbox for mark summary characteristics. By applying compositional data analysis principles \citep{10.5555/17272} to marked spatial point processes, this work goes beyond existing methods, establishing novel summary characteristics for complex mark scenarios. To our knowledge, composition-valued marked point processes are examined here for the first time.

Various summary characteristics are standard tools to characterize both the properties of points and the interrelations between marks (and marks and points) \citep[][]{StoyanStoyan1994,  Illian2008, Chiu2013}. 
Of all established methods, functional summary characteristics, in which the characteristic is defined as a function of the distance between points $r$, play a particularly important role. They are used to investigate the marked point pattern, to determine a suitable model and to evaluate the goodness-of-fit of such models \citep[e.g.][]{Illian2008, mari1}. 

Despite the growing availability of highly demanding spatial marked point process scenarios,
the literature focused, so far, almost exclusively on the analysis of scalar-valued attributes \citep[see][for a general treatment]{Baddeley2010}. For integer-valued, i.e. qualitative, marks, where the points are assigned to exactly one out of $k\geq 2$ distinct types, so-called \textit{cross}- and \textit{dot}-type extensions of classic nearest-neighbour and pairwise distance based summary characteristics exist \citep{Lotwick1982, HarknessIsham1983, DIGGLE1986115, VanLieshoutBaddeley1999}. For real-valued marks, the aim has mainly been  to quantify the average association or variation among the mark values for pairs of distinct points at a distance $r$ apart. This includes characteristics which highlight the spatial mark-to-mark \citep{Stoyan1984,Stoyan1987,StoyanStoyan1994, Stoyan2000,Schlather2001}
and point-to-mark associations \citep{Schlather2004, Guan2006, Guan2007a, HO20081194, ZHANG201420}, (real-valued) mark weighted versions of classic point summary characteristics \citep{Penttinen1992, Lieshout2006:realval:J} and frequency domain approaches \citep{EckardtISR, Eckardt2018partial}. 
Further, \cite{Stoyan1987} considered the analysis of two distinct real-valued marks, and \cite{Penttinen1992}, \cite{WiegandMoloney2013} and \cite{EckardtISR,Eckardt2018partial} discussed methods for mixtures of integer- and real-valued marks. 
Further, function-valued marks \citep{comas2008METMA,Comas2011,Comas2013,Ghorbani2020} and   generalizations to multivariate function-valued marks  \citep{Eckardt2023MultiFunctionMarks}
have been investigated \citep[see][for a general review]{Eckardt:Moradi:currrent, Eckardt2024Rejoinder}. However, there are no methods available for the joint analysis of (constrained) vector-valued marks.

We extend the considered mark quantities. This paper aims to characterise the spatial association for pairs of points with $D$-part composition-valued marks. 
Instead of investigating the component parts 
separately, 
the underlying idea is to treat the observed marks as part of a total, and to transfer the main principles of compositional data analysis to marked spatial point processes. Composition-valued geostatistical and areal data has already received much attention, including various structural analysis and kriging approaches \citep{pawlowsky1986raumliche,PawlowskyOrea2004,Tologado2006PhD} and areal regression models \citep{5bcbae68-b00d-3abd-8e59-25b71c41d7df, Huang2021}. However, no extensions to marked point processes with composition-valued marks exist so far. As our main contribution, we introduce a general framework of composition-valued marks and define appropriate mark characteristics for them.

The remainder of the paper is structured as follows. Section \ref{sec:2ndsummary:statusquo} summarizes the present methodological toolbox for real-valued marks we build on. Section \ref{sec:coda} introduces composition-valued marks, suitable mark transformations and first-order tools (Section \ref{sec:CompMarks}-\ref{sec:1storder}), different mark summary characteristics (Sections \ref{sec:SummaryChrCoDa}-\ref{sec:CodaGlobal}), extensions to composition-valued marks with total information (Section \ref{sec:CodaTotal}) and introduces estimation for the new methods (Section \ref{sec:estimation}).
Section \ref{sec:testing} considers testing of the basic random labeling hypothesis for composition-valued marks. Applications of the proposed methods to a Finnish tree pattern and Spanish business sector data are provided in Section \ref{sec:appl}. The paper closes with a conclusion in Section \ref{sec:conclusio}.



\section{Summary characteristics for real-valued point attributes}\label{sec:2ndsummary:statusquo}

As compositions can be seen as a constrained vector of real-valued marks, the following discussion mostly  restricts to methods for the real-valued marks scenario, which are to be extended for composition-valued marks in Section \ref{sec:coda}. 



\subsection{Preliminaries}

Let $X=\lbrace x_i, m(x_i)\rbrace_{i=1,\ldots, n}$ denote a marked spatial point process on $\R^2\times \mathds{M}$ with points $x_i$ in a two-dimensional Euclidean space and associated marks $m(x_i)$ living in a mark space $\mathds{M}$. The observed point pattern and the related  unmarked point process will be denoted by $\mathbf{x}$
and $\X
$, respectively. In what follows, $X$ is assumed to be simple, where simplicity means that multiple coincident points do not occur. In general, we assume $\M$ to be a 
Polish space equipped with a $\sigma$-algebra $\mathcal{M}$ and an appropriate reference measure $\varpi$. The mark distribution will be denoted by $M$. The Borel $\sigma$-algebra of $\R^2$ is denoted by $\mathcal{B}$. For $X$, the expected number $N(\cdot)$ of points in $B \in \mathcal{B}$ with marks in $L\in \mathcal{M}$ is 
\[
\Lambda(B\times L) = \e(N(B\times L)).
\]
Further,  $X$ is said to be stationary if $\lbrace x_i, m(x_i)\rbrace \overset{d}{=}\lbrace x_i+s, m(x_i)\rbrace$ for all $s \in \R^2$ and isotropic if $\lbrace x_i, m(x_i) \rbrace \overset{d}{=}\lbrace \mathfrak{r}x_i, m(x_i)\rbrace$ for any rotation $\mathfrak{r}\in \R^2$, $\overset{d}{=}$ denoting equality in distribution. If $X$ is stationary and isotropic, $X$ is called motion-invariant. For stationary $X$, $\Lambda(B\times L)$ simplifies to 
\[
\Lambda(B \times L)= \lambda\nu(B)M(L),
\]
where $\lambda$ is the intensity of $\X$, and $\nu(\cdot)$ is the Lebesgue measure. 
If $\mathds{M}=\mathds{R}$, $M$ is completely determined by the mark distribution function $F_M(m)=M(\left(-\infty, m \right])$ for $-\infty\leq m\leq \infty$, where  
\[
F_M(m)=\int_{-\infty}^{m} f_M(\Tilde{m})\de \Tilde{m},
\]
with $f_M(\cdot)$ denoting the mark density function if it  exists. Here, $f_M(L)\coloneq\int_L f_M(l)\de l$ can be interpreted as the probability that at an arbitrarily chosen point the mark is in $L$. Useful quantities of the mark distribution are the mean mark $\mu_M$ 
and the mark variance $\sigma^2_M$.
Finally, the second-order factorial moment measure of $\X$ plays an important role in the second-order summary characteristics. It is defined as 
\begin{eqnarray}\label{eq:factorial:measure}
\alpha^{(2)}(B_1\times B_2)&=\e\left[\sum^{\neq}_{x_1,x_2~\in \X} \mathds{1}_{B_1}(x_1)\mathds{1}_{B_2}(x_2)\right]\\
&=\int_{B_1}\int_{B_2}\varrho^{(2)}(x_1, x_2)\de x_1 \de x_2\nonumber,
\end{eqnarray}
where the symbol $\sum^{\neq}$ denotes the sum over distinct pairs of points, $\mathds{1}_B(x)$ an indicator function of whether $x\in B$ and $\varrho^{(2)}$ is the second-order product density. Heuristically, for any $x_1, x_2\in \mathds{R}^2$, $\varrho^{(2)}(x_1, x_2)\de x_1 \de x_2$ can be interpreted as the probability of observing exactly one point in each of the infinitesimal areas $\de x_1$ and $\de x_2$.

\subsection{Functional summary characteristics for real-valued marks}\label{sec:secondorder:realvalues}

Within the last decades various mark characteristics were introduced. These characteristics either describe the average pairwise variation or association of the marks at two point locations as a function of the interpoint distance $r$. 
Prominent cases include Stoyan's mark correlation function \citep{StoyanStoyan1994}, the mark variogram \citep{cressie93}, 
 the mark covariance function \citep{DBLP:journals/eik/Stoyan84}, Isham's mark correlation function \citep{isham1985marked}, and Schlather's \citep{Schlather2004} and Shimatani's \citep{Shimatani:MoranI} $I$ functions. 
All these characteristics  are defined exclusively  for stationary point processes. 
These characteristics are conditional quantities \citep[in a Palm sense, see][]{Chiu2013}, i.e.\ conditional on that there are indeed points at location $\circ$ and $\mathbf{r}$ in $\X$. 
They are commonly constructed by taking the conditional expectation $\e_{\circ,r}$ of a so-called \textit{test function} $\tf_f: \mathds{M}\times\mathds{M} \to \R^+$, which takes the marks $m(\circ)$ and $m(\mathbf{r})$ at the origin $\circ$ and any alternative points at distance $\Vert\mathbf{r}\Vert=r>0$ from $\circ$ as its arguments \citep{PenttinenandStoyan1989}. Without imposing any invariance assumptions, the expectation is formally defined with respect to the joint distribution of the marks $m(x_1)$ and $m(x_2)$ for any two points $x_1, x_2\in \mathds{R}^2$ under the condition that there are indeed points at locations $x_1$ and $x_2$ in $\Breve{X}$, i.e.\ the so-called two-point mark distribution $M_{x_1,x_2}(\de m(x_1) \de m(x_2))$. For motion-invariant $X$, $M_{x_1,x_2}$ depends on the points only through the Euclidean distance $\Vert x_1 - x_2 \Vert=r$ and can be written as $M_r$. For $L_1,L_2\in\mathcal{M}$, $M_r(L_1\times L_2)$ corresponds to the probability of having $m(\circ)\in L_1$ and $m(\mathbf{r})\in L_2$ under the condition that there are indeed points at $\circ$ and $\mathbf{r}$ at a distance $r$ from each other in $\X$. Note that $M_r(L_1\times L_2)=\varrho^{(2)}(r, L_1, L_2)/\varrho^{(2)}(r)$  simplifies to  $(\varrho^{(2)}(r)M(L_1)M(L_2))/\varrho^{(2)}(r)=M(L_1)M(L_2)$ under independent marks for all $L_1$ and $L_2$ in $\mathcal{M}$, where 
$\varrho^{(2)}(r, L_1, L_2)$ is the second-order product density of 
\[
\alpha^{(2)}_m(B_1\times L_1 \times B_2 \times L_2) =\int_{B_1 \times B_2} M_{r}(L_1 \times L_2)\alpha^{(2)}\lbrace d(x_1,x_2)\rbrace
\]
and $\varrho^{(2)}(r)$ is the second-order product density as in \eqref{eq:factorial:measure} for $x_1, x_2$ at distance $r$ \citep{PenttinenandStoyan1989}.

Using the above notation, functional mark characteristics are  constructed as follows. Denoting by $\nabla_{\tf_f}(r)=\e_{\circ,r}\left[\tf_f(m(\circ), m(\mathbf{r}))\right]$ and writing  $\nabla_{\tf_f}=\nabla_{\tf_f}(\infty)$ for the expected value of the chosen test function $\tf_f$ at very large distances, i.e.\ when the marks are expected to be independent, the specific mark characteristic itself is determined by the specific choice of the test function $\tf_f$ \citep[see][]{Illian2008}. Formally, $\nabla_{\tf_f}(r)$ can be expressed as the ratio of two product density functions, 
\begin{equation}
     \nabla_{\tf_f}(r)=\varrho^{(2)}_{\tf_f}(r)\left/\varrho^{(2)}(r)\right., 
\end{equation}
i.e.\ the densities
of the $\tf_f$-factorial moment measure $\alpha_{\tf_f}^{(2)}(B_1 \times B_2)$,
\begin{eqnarray}
\alpha_{\tf_f}^{(2)}(B_1 \times B_2)
&=
\e
\left[
\sum^{\neq}_{\substack{(x_1,m(x_1)),\\ (x_2,m(x_2))~\in X}}
\tf_f(m(x_1),m(x_2))
\mathds{1}_{B_1}(x_1)
\mathds{1}_{B_2}(x_2)
\right],
&\\
\nonumber
\end{eqnarray}
and of the factorial moment measures $\alpha^{(2)}(B_1\times B_2)$ of \eqref{eq:factorial:measure}, 
respectively, where $B_1, B_2$ are sets in $\mathcal{B}$. When the distance $r$ tends to infinity, the marks are assumed to be independent, $\nabla_{\tf_f}(r)$ becomes independent of the points and simplifies to 
\begin{eqnarray}\label{eq:ExpectedIndMarks}
    ´\nabla_{\tf_f}=
    \int_\M\int_\M\tf_f(m_1,m_2) F_M(\de m_1) F_M(\de m_2).
\end{eqnarray}
An overview of the most prominent specifications for $\tf_f$ is presented in Table \ref{tab:testfunctions}. 
\begin{table}
    \centering
    \begin{tabular}{lllll}
    \hline
     Name   & Test  & Normalising & Notation & Notation \\
    for $\tf_f$   & function $\tf_f$ &    factor $\nabla_{\tf_f}$ & for $\nabla_{\tf_f}(r)$ & for $\kappa_{\tf_f}(r)$\\
    \hline
$\tf_1$ & $m(\circ)m(\mathbf{r})$ & $\mu_M^2$ & $\tau_{mm}(r)$ & $\kappa_{mm}(r)$ \\
$\tf_2$ & $m(\circ)$  & $\mu_M$ & $\tau_{m\bullet}(r)$ & $\kappa_{m\bullet}(r)$ \\ 
      $ \tf_3$ & $m(\mathbf{r})$ &  $\mu_M$ & $\tau_{\bullet m}(r)$ & $\kappa_{\bullet m}(r)$ \\ 
  $\tf_4$ & $0.5(m(\circ)- m(\mathbf{r}))^2$  & $\sigma^2_M$
  & $\gamma_{mm}(r)$ & $\gamma^{\mathrm{n}}_{mm}(r)$\\
    $\tf_5$ & $(m(\circ)-\mu_M)(m(\mathbf{r})-\mu_M)$   &$\sigma^2_M$  &  $\iota^{\mathrm{Shi}}_{mm}(r)$ &  $I^{\mathrm{Shi}}_{mm}(r)$\\
    $\tf_6$ & $(m(\circ)-\mu_M(\mathbf{r}))(m(\mathbf{r})-\mu_M(\mathbf{r}))$  & $\sigma^2_M$ &  $\iota_{mm}^{\mathrm{Sch}}(r)$ &  $I^{\mathrm{Sch}}_{mm}(r)$ \\
   \hline
    \end{tabular}
    \caption{Overview of prominent test function specifications with $\mu_M$ and $\sigma_M^2$ denoting the unconditional mark mean and mark variance, $\mu_M(\mathbf{r})$ the mean for the second point at location $\mathbf{r}$ under the condition that there are points at $\circ$ and $\mathbf{r}$.}
    \label{tab:testfunctions}
\end{table}

Evaluating the test functions of Table \ref{tab:testfunctions} immediately yields different functional mark summary characteristics, which we briefly discuss next.
With the exception of the last two test function, the normalising factor $\nabla_{\tf_f}$ (3rd column) corresponds to the expected valued of the test function  if $r\to \infty$ and follows directly from the solution of \eqref{eq:ExpectedIndMarks} with $\tf_f(m_1,m_2)$ replaced by the specific test function (as given in the 2nd column of Table \ref{tab:testfunctions}) \citep[see][]{Illian2008}. For $\tf_5$ and $\tf_6$, however, $\nabla_{\tf_f}$ is set to $\sigma_M^2$ in close analogy to Moran's $I$ \citep{moran}. 
Instead of $\nabla_{\tf_f}(r)$, it is sometimes preferable to compute the $\tf_f$-correlation function
\citep{PenttinenandStoyan1989}, 
\begin{equation}\label{eq:tfcorr}
     \kappa_{\tf_f}(r)=\frac{\nabla_{\tf_f}(r)}{\nabla_{\tf_f}}  = \left.\frac{\varrho^{(2)}_{\tf_f}(r)}{\varrho^{(2)}(r)}\right/ \nabla_{\tf_f},
\end{equation}
which normalizes the conditional expectation of the test function $\nabla_{\tf_f}(r)$  by its expectation $\nabla_{\tf_f}$ for $r\to\infty$ 
such that $\kappa_{\tf_f}(r)=1$ for all $r$ 
under mark independence (or by $\sigma_M^2$ in the last two cases). For completeness, Table \ref{tab:testfunctions} covers both the unnormalized ($\nabla_{\tf_f}(r)$) and related normalized ($\kappa_{\tf_f}(r)$) expressions.  

A classic summary is the conditional mean product of marks of two points being a distance $r$ apart, $\tau_{mm}(r)$, and its scaled version $\kappa_{mm}(r)$ often termed the mark correlation function  \citep{StoyanStoyan1994}.
If large (resp. small) marks systematically co-occur at interpoint distance $r$, their pairwise products will also be large (resp. small) and deviate from the independent mark assumption, i.e.\ the mark mean squared. 
Both \textbf{r}-mark functions $\tau_{m\bullet}$ and $\tau_{\bullet m}$ and the related \textbf{r}-mark correlation functions $\kappa_{m\bullet}$ and $\kappa_{\bullet m}$ can be interpreted as the conditional expectation of the mark of a point given that there is another point with distance $r$. This expectation often deviates from $\mu_M$, the expectation under independent marks, when the marks are dependent on the existence of other points \citep{Schlather2004,  myllymaki_deviation_2015}.
The mark variogram $\gamma_{mm}(r)$ is a measure of the average dispersion \citep{cressie93}. 
It helps to detect situations where the marks of points close together tend to be more similar (or different) than expected under mark independence.
Shimantani's \citep{Shimatani:MoranI} and Schlather's \citep{Schlather2001} $I$-functions $\iota_{mm}^{\mathrm{Shi}}$ and $\iota_{mm}^{\mathrm{Sch}}$ and their normalized versions $I_{mm}^{\mathrm{Shi}}$ and $I_{mm}^{\mathrm{Shi}}$ can be seen as adaptations of Moran's $I$ \citep{moran} to spatial point processes. They are helpful for identifying potential spatial autocorrelation among the marks. Although these test functions are similar in spirit, Schlather applies a centering not by the unconditional mean but by the 
conditional mean $\e_{\circ,r}\left[m(\mathbf{r})\right]=\mu_M(\mathbf{r})$ of the second point.
We note that the above construction principle through differently specified test functions can also be applied to define so-called  nearest-neighbour correlation indices, i.e.\ numerical mark summary characteristics, where instead of $m(\circ)$ and $m(\mathbf{r})$ the test functions only consider the marks at the origin and its nearest-neighbouring point(s) \citep[see][for a detailed discussion]{StoyanStoyan1994}. 
 
\subsection{Mark-weighted summary characteristics for real-valued marks}\label{sec:nn:index}

Apart from the above summary characteristics, some authors propose to use the test function as a weight for classic second-order spatial point process characteristics. Initially proposed by  \cite{Penttinen1992},  the mark-weighted  $K_{\tf_f}$ function, which is a generalisation of Ripley's $K$-function \citep{Ripley1976} to real-valued marks, is defined as
\begin{equation}\label{eq:Kmarked}
   K_{\tf_f}(r)=\frac{\e_{\circ,r}\left[\sum_{(x_i,m(x_i))\in X} \tf_f(m(\circ), m(x_i))\mathds{1}_{b(\circ,r)}\lbrace x_i\rbrace\right]}{\lambda \nabla_{\tf_f}}.  
\end{equation} 
Here, $b(\circ, r)$ is a disc of radius $r$ centered at the origin, $\lambda$ is the intensity of $\X$, and $\tf_f(\cdot)$ is any test function as presented in Table \ref{tab:testfunctions}. 
Again, the precise interpretation of the $K_{\tf_f}$ 
function depends on the specific test function under study. For $\tf_f=\tf_1$, $\nabla_{\tf_f}$ equals $\mu_M^2$ and $K_{\tf_f}$ is denoted by $K_{mm}$, 
\begin{equation}\label{eq:Kmarked:mm}
   K_{mm}(r)=\frac{\e_{\circ,r}\left[\sum_{(x_i,m(x_i))\in X} m(\circ)\cdot m(x_i)\mathds{1}_{b(\circ,r)}\lbrace x_i\rbrace\right]}{\lambda\mu_M^2}.
\end{equation} 
Recalling the definition of Ripley's $K$ function, $K_{mm}$ can be interpreted as the expected number of further points within a distance $r$  weighted by the pairwise product of marks. If the average product of marks coincides with the mark mean squared, $\nabla_{\tf_1}(r)$ equals  $\mu_M^2$ and $K_{mm}$ reduces to $K$. However, for dependent marks, i.e. when $\nabla_{\tf_1}(r)>\mu_M^2$ (resp. $\nabla_{\tf_1}(r)<\mu_M^2$) for some $r$, $K_{mm}(r)>K(r)$  (resp. $K_{mm}(r)<K(r)$). 
We note that, for nicer visualization, it is often preferable to use $L_{\tf_f}(r)=\sqrt{K_{\tf_f}(r)/\pi}$, a variance stabilising and centered version of $K_{\tf_f}$
instead of $K_{\tf_f}$ to control for the strict monotonic behaviour of the $K$-function with respect to the distance $r$. 

\section{Composition-valued marked point processes}\label{sec:coda}

\subsection{Composition-valued marks}\label{sec:CompMarks}

To extend spatial point processes to composition-valued marks, let $\lbrace (x_i, \cb(x_i) )\rbrace^n_{i=1}$ denote a set of $n$ points $x_i\in\mathds{W}\subset\mathds{R}^2$ with associated marks  $\cb(x_i)=(c_{1}(x_i),\ldots, c_{D}(x_i))^{\top}$ living in a $D$-part simplex $\si^D\subset\mathds{R}^D$,  where for some $\mathfrak{w}\in\mathds{R}$
\[
\mathds{S}^{D}=\left\{\cb =(c_{1},c_{2},\dots ,c_{D})^{\top}\in \mathds {R}^{D}\,\left|\,c_{j}\geq 0,j=1,2,\dots ,D;\sum_{j=1}^{D}c_{j}=\mathfrak{w} \right.\right\}. 
\]
Common choices of $\mathfrak{w}$ include $\mathfrak{w}=1$ and $\mathfrak{w}=100$ (per cent) depending on the marks at hand. 
Thus, the mark at each location is a composition of $D$ non-negative parts summing to a constant and we call any such mark composition-valued. That is, composition-valued marked spatial point processes are a particular type of spatial compositional data \citep{10.5555/17272}, where each mark $\cb(x_i)$ quantitatively describes the relative importance of the individual parts with respect to a given total. 
We note that composition-valued marks could have two potential origins: they might (i) directly arise from the data collection 
or (ii) be constructed from non-simple spatial point process scenarios with integer-valued marks, by summarizing the absolute numbers $\tilde{c}_j,\ j=1,\ldots, D$, at a point location a-posteriori into relative numbers for distinct categories. Generally, any such absolute information can always be transformed into a composition-valued mark by dividing each part by the sum over all components, i.e.\ by applying the closure operation $\cb=\cls(\tilde{\cb})=\left(\tilde{c}_1 / \sum_{j=1}^D \tilde{c}_j,\ldots,\tilde{c}_D / \sum_{j=1}^D \tilde{c}_j\right)^\top$. Consequently, both the absolute and the relative, composition-valued marks can be analysed depending on the question of interest, providing two different views of the marks as discussed below in Section \ref{sec:CodaTotal}. 


The $\si^D$ space of compositions can be equipped with a finite $(D-1)$ dimensional Euclidean vector space structure, i.e.\ the  Aitchison geometry, with the perturbation $\cb\oplus\mathbf{c'}=\cls(c_1c'_1,c_2c'_2,\ldots, c_Dc'_D)$, a commutative group operation on the simplex with neutral element $\mathbf{n}= \cls(1, 1,\ldots, 1)$ and inverse operation $\cb\ominus\cb'=\cb\oplus((-1)\odot\cb')$, and the powering  operation   $\xi\odot\cb=\cls(c_1^{\xi},c_2^{\xi},\ldots, c_D^{\xi})$,  where $\mathbf{c,c'}\in\mathds{S}^{D}$ and $\xi\in\mathds{R}$  \citep{MR1873662}. Additionally, the inner product 
is defined as
\begin{equation}\label{eq:AitInner}
   \langle\mathbf{c,c'}\rangle_A=\frac{1}{2D}\sum^{D}_{l=1}\sum^D_{j=1} \log\left(\frac{c_l}{c_j}\right)\log\left(\frac{c'_l}{c'_j}\right) 
\end{equation}
yielding the norm $\Vert\cb \Vert_A=\sqrt{\langle\mathbf{c,c}\rangle_A}$  and the associated distance 
$d_A(\mathbf{c,c'})=\Vert\cb\ominus\mathbf{c'} \Vert_A$. Note that the composition-valued marks can be transformed to real-valued coordinates through a map function $\psi:\mathds{S}^D\to \mathds{R}^{\tilde{D}}, \cb\mapsto \psi(\cb)$ where ${\tilde{D}}$ is determined by the particular choice of $\psi$. For particular useful choices, an isometric isomorphism exists between the $\si^D$ and the $\mathds{R}^{\Tilde{D}}$  \citep{BillheimerEtAl2001, PawlowskyGlahn2001}, which allows expressing the composition-valued marks in real coordinates such that their relations are correspondent to those in the Aitchison geometry. Below in Section \ref{sec:transformations}, we give alternatives for the transformation $\psi$. After transformation, statistical analysis methods can be performed in $\mathds{R}^{\tilde{D}}$ \citep{doi:https://doi.org/10.1002/9781119976462.ch3}. 


\subsection{Transformations}\label{sec:transformations}

Writing $\psi_j(\cb)$ for the $j$-th element of $\psi(\cb)=(\psi_1(\cb),\ldots, \psi_{\tilde{D}}(\cb))^\top$, 
different coordinate representations can be derived \citep{doi:10.1002/9781119976462}. 
Apart from the  \textit{log-ratio} (lr) transformation early specifications of $\psi$ 
include the \textit{additive log-ratio} (alr) transformation  \citep{10.1093/biomet/67.2.261} where $\psi_j(\mathbf{c})=\log(c_j/c_D),\ j= 1,\ldots,{\tilde{D}}=D-1$ i.e.\ the log-ratios relative to the $D$-th component, and the \textit{centered log-ratio} (clr)  \citep{10.1093/biomet/70.1.57} transformation $\psi_j(\mathbf{c})=\log(c_j/g(\cb)),\ j=1,\ldots, \Tilde{D}=D$, i.e.\ the log-ratios relative to the geometric mean $g(\cb)=(\prod_{j=1}^D c_j)^{1/D}$. While the  alr transformation yields an isomorphic but non-isometric 
relation between the two spaces, i.e.\ it does not preserve distances, the clr transformation results in an isometric isomorphism by mapping the marks from the simplex to a hyperplane $\mathds{H}\subset\mathds{R}^{D}$ that is orthogonal to the  vector of ones. However, the clr imposes a  sum-to-zero constraint on the transformed marks, which can make analysis more difficult e.g.\ due to degenerated distributions and singular covariance matrices.

We observe that neither the alr nor the clr transformation can be directly linked to an orthonormal coordinate system on the simplex. However, an orthonormal basis $(\mathbf{e}_1, \mathbf{e}_2,\ldots,\mathbf{e}_{D-1})$ on the simplex $\mathds{S}^D$  with respect to the inner product can be derived using the Gram-Schmidt procedure giving 
\[
\cb=\bigoplus_{j=1}^{D-1}\langle\cb,\mathbf{e}_j\rangle_A\odot\mathbf{e}_j.
\]
This coordinate representation corresponds to the \textit{isometric log-ratio} ($\ilr$) transformation, a class of orthonormal coordinate representations, defined by 
\[
\ilr(\cb) = \left(\langle\cb,\mathbf{e}_1\rangle_A,\langle\cb,\mathbf{e}_2\rangle_A,\ldots,\langle\cb,\mathbf{e}_{D-1}\rangle_A\right)
\]
and establishes an isometric isomorphism through the map between $\mathds{S}^D$ and $\mathds{R}^{D-1}$. 
Further, the $\ilr$  transformation is related to the clr and log transformations through $\ilr(\cb)=\clr(\cb)\mathbf{H}_\mathrm{D}^\top=\log(\cb)\mathbf{H}_\mathrm{D}^\top$ where 
$\mathbf{H}_\mathrm{D}$ denotes a $((D-1)\times D)$-dimensional Helmert matrix with rows $\mathbf{h}_j=\clr(\mathbf{e}_j), j=1,\ldots,D-1$  satisfying $\mathbf{H}_\mathrm{D}\mathbf{H}_\mathrm{D}^{\top}=\mathbf{I}_{\mathrm{D}-1}$ and $\mathbf{H}_\mathrm{D}^{\top}\mathbf{H}_\mathrm{D}=\mathbf{G}_\mathrm{D}$ and $\mathbf{G}_\mathrm{D}$ is the $D$-dimensional centering matrix $\mathbf{G}_\mathrm{D}=\mathbf{I}_{\mathrm{D}}-D^{-1}\mathds{1}_\mathrm{D}\mathds{1}^{\top}_\mathrm{D}=\mathbf{G}_\mathrm{D}^2$, $\mathbf{I}_{\mathrm{D}}$ is the identity matrix of dimension $(D\times D)$, and $\mathds{1}_\mathrm{D}$ a $(D\times 1)$ vector of ones.    
Due to the isometric isomorphism established between the  Aitchison and the Euclidean geometry by the $\clr$ and the $\ilr$ transformations, the  Aitchison inner product, distances and metrics coincide with their Euclidean counterparts on the transformed quantities, such that for any $\mathbf{c,c'}\in\mathds{S}^D$
\begin{align}\label{eq:AitDist}
    d_A(\cb,\mathbf{c'})&=d_E(\clr(\cb),\clr(\mathbf{c'}))=d_E(\ilr(\cb),\ilr(\mathbf{c'}))
\end{align}
with $d_E$ the Euclidean distance. 
The $\ilr$ transformation is equivalent to the logit-function used in logistic regression if $D=2$. When $D>2$, infinitely many  orthonormal basis systems exist and the concrete choice has a crucial impact on the interpretation of the projected data. Particular choices of an orthonormal basis are coordinate representations using (i) \textit{balances} \citep{Egozcue2005:Balances} in which each \textit{balancing element} can be interpreted as the normalised log-ratio of the geometric means (centers) of two groups, and (ii) \textit{pivot coordinates} \citep{Fiserova2011, Hron2017}.
Balances originate from a \textit{sequential binary partition} method, which involves dividing the composition into two parts. The $j$-th ilr coefficient using pivot coordinates can be expressed as  
\begin{equation*}
     \ilr_j(\cb)=\sqrt{\left(\frac{D-j}{D-j+1}\right)}\log\Bigg\{\frac{c_j}{\sqrt[D-j]{\prod_{k=j+1}^Dc_k}}\Bigg\}, \ j=1,\ldots, D-1
\end{equation*}
\citep{Fiserova2011}. The initial ilr coefficient is akin to the first clr coefficient, scaled by $\sqrt{D/(D-1)}$, and is easily interpreted as the log-ratio of the respective component to the geometric mean. In contrast, interpreting subsequent coefficients is more complex. To address this, the literature proposes generalised pivot coordinates using permuted compositions and symmetric pivot coordinates \citep[see e.g.][]{sym:pivot, Hron2021}.


While the above transformations allow for a representation of the composition-valued marks in coordinates in  a Euclidean space, the underlying log-operations are undefined for zero values. In what follows, we assume that the proportions for all $D$-parts are non-zero. However, as zeros might be present in some marked point process scenarios when potentially not all components are observed at each location, we also provide a treatment of different transformations in the presence of structural zeros, see Section 1 in the supplementary material.

\subsection{First-order tools for composition-valued marks}\label{sec:1storder}

We first review first-order mark characteristics for the composition-valued marks.
General characteristics for a  strictly-positive sample $\cb_1, \dots, \cb_n$ of  compositions  commonly applied in the literature include the geometric center, i.e. the closed geometric mean, 
\begin{equation}\label{eq:centerCoDa}
 \cen(\cb)=\frac{1}{n}\odot\bigoplus^n_{j=1}\cb_j=\clrinv\left(\frac{1}{n}\sum^n_{j=1}\clr(\cb_j)\right),   
\end{equation}
where $\clrinv(\Tilde{\cb})=\cls(\exp(\Tilde{\cb}))$.
It serves as the mean composition-valued mark.
Further, the variation of the composition-valued marks can be described through the variation matrix $\mathbf{T}$ 
with elements $t_{jl}=\var\left[\log(c_j/c_l)\right], j,l=1,\ldots, D$,
and the total, i.e.\ metric, variance 
\[
\mvar\left[\cb\right]=\frac{1}{2D}\sum_{j,l=1}^D t_{jl}=\frac{1}{n-1}\sum_{j=1}^n d_A^2(\cb_j,\cen(\cb)),
\]
serving as a global measure of dispersion 
\citep{10.5555/17272, PawlowskyGlahn2001}. Instead of the variation matrix, a normalised variation matrix $\NT=0.5\cdot\mathbf{T}$ can be used. 
 
In the presence of zero components, alternative measures of centrality include the spatial median \citep{10.2307/2345619}, the  graph median \citep{Sharp2006} 
and the Fr{\'e}chet mean of the $\alpha$-transformed composition \citep{alpha:2011}.

\subsection{Componentwise summary characteristics for composition-valued marks}\label{sec:SummaryChrCoDa}

We now define novel componentwise characteristics for composition-valued marks analogously to the mark characteristics of Section \ref{sec:secondorder:realvalues}. These are useful to look at the spatial behaviour  of all components individually, while we propose characteristics for the whole composition in the next subsection. We here explicitly focus on the case where each point is augmented by exactly one composition. Extensions to multivariate settings are outlined in Section \ref{sec:CodaMulti} of the Supplementary material.
Recall that $\cb(\circ)$ and $\cb(\mathbf{r})$ denote the composition-valued marks for a pair of points at locations $\circ$ and $\mathbf{r}$ at distance $\Vert\mathbf{r}\Vert=r$, and
$\psi(\cb)$ is the transformed composition-valued mark with $j$th element or component $\psi_j(\cb)$, where $\psi:\si^D\mapsto\mathds{R}^{\tilde{D}}$.
We then define
$$
\nabla_{\tf_f}^{\psi,jl}(r)=\e_{\circ, r}\left[\tf_f^{\psi, jl}(\psi_j(\cb(\circ)), \psi_l(\cb(\mathbf{r})))\right],
$$
where $\tf_f^{\psi, jl}$ denotes a test function specific to the transformed composition-valued marks. The test functions of Table \ref{tab:testfunctions} can be employed as $\tf_f^{\psi, jl}$; for clarity, they are re-expressed for the transformed marks $\psi(\cb)$ in Table \ref{tab:testfunctions:CoDa}.
Further, we let $\nabla_{\tf_f}^{\psi,jl}$ stand for the limiting case of $\nabla_{\tf_f}^{\psi,jl}(r)$ when $r\to \infty$, i.e.\ 
\begin{eqnarray}\label{eq:ExpectedIndMarksPsi}
    \nabla_{\tf_f}^{\psi,jl}=
    \int_{\R^{\Tilde{D}}}\int_{\R^{\Tilde{D}}}\tf_f^{\psi, jl}(\psi_j(c_1),\psi_l(c_2)) \varpi(\de \psi_j(c_1))\varpi(\de \psi_l(c_2)),
\end{eqnarray}
and define the $\tf_f^{\psi,jl}$-correlation function $\kappa^{\psi,jl}_{\tf_f}(r)$ as
\begin{equation}\label{eq:ktf:fct:psi}
    \kappa^{\psi,jl}_{\tf_f}(r) = \nabla_{\tf_f}^{\psi,jl}(r)\left/\nabla_{\tf_f}^{\psi,jl}\right.,
\end{equation}
where \eqref{eq:ExpectedIndMarksPsi} is again replaced by the corresponding variance in the case of the last two test functions.
The choice of the test function $\tf_f^{\psi, jl}$ determines the focus of the analysis, as the different characteristics highlight different properties of marks (see Section \ref{sec:secondorder:realvalues}).

\begin{table}
    \centering
    \begin{tabular}{lllll}
    \hline
     Name   & Test  & Normalising & Notation & Notation \\
    for $\tf_f^{\psi, jl}$   & function $\tf_f^{\psi, jl}$ &    factor $\nabla^{\psi, jl}_{\tf_f}$ & for $\nabla^{\psi, jl}_{\tf_f}(r)$ & for $\kappa^{\psi, jl}_{\tf_f}(r)$\\
    \hline
$\tf_1^{\psi, jl}$ & $\psi_j(\mathbf{c}(\circ))\psi_l(\mathbf{c}(\mathbf{r}))$ & $\mu_{j}^{\psi} \mu_{l}^{\psi}$ & $\tau^{\psi}_{jl}(r)$ & $\kappa^{\psi}_{jl}(r)$ \\
$\tf_2^{\psi, jl}$ & $\psi_j(\mathbf{c}(\circ))$       & $ \mu^{\psi}_{j}$ & $\tau^{\psi}_{j\bullet}(r)$ & $\kappa^{\psi}_{j\bullet}(r)$ \\ 
      $ \tf_3^{\psi, jl}$ & $\psi_l(\mathbf{c}(\mathbf{r}))$ & $ \mu^{\psi}_{l}$ & $\tau^{\psi}_{\bullet l}(r)$ & $\kappa^{\psi}_{\bullet l}(r)$ \\ 
  $\tf_4^{\psi, jl}$ & $0.5(\psi_j(\mathbf{c}(\circ))- \psi_l(\mathbf{c}(\mathbf{r})))^2$ & $\zeta_{jl}^{\psi}$
  & $\gamma^{\psi}_{jl}(r)$ & $\gamma^{\psi,\mathrm{n}}_{jl}(r)$\\
    $\tf_5^{\psi, jl}$ & $(\psi_j(\mathbf{c}(\circ))-\mu^{\psi}_{j})(\psi_l(\mathbf{c}(\mathbf{r}))-\mu^{\psi}_{l})$ &$\sigma_{jl}^{\psi}$  &  $\iota^{\psi,\mathrm{Shi}}_{jl}(r)$ &  $I^{\psi, \mathrm{Shi}}_{jl}(r)$\\
    $\tf_6^{\psi, jl}$ & $(\psi_j(\mathbf{c}(\circ))-\mu^{\psi}_{j}(\mathbf{r}))(\psi_l(\mathbf{c}(\mathbf{r}))-\mu^{\psi}_{l}(\mathbf{r}))$ & $\sigma_{jl}^{\psi}$ &  $\iota^{\psi, \mathrm{Sch}}_{jl}(r)$ &  $I^{\psi, \mathrm{Sch}}_{jl}(r)$ \\
   \hline
    \end{tabular}
    \caption{Specifications for the componentwise test function $\tf_f^{\psi,jl}(\psi_j(\mathbf{c}(\circ)),\psi_l(\cb(\mathbf{r})))$  with $\mu^{\psi}_{j}$ denoting the unconditional mark mean and 
    $\mu^{\psi}_{j}(\mathbf{r})$ the conditional mark mean of the $j$-th element of the $\psi$-transformed composition-valued marks,
    where $\zeta_{jl}^{\psi}= 0.5\left[ 
    \sigma^{\psi}_{jj} + 
     \sigma^{\psi}_{ll} + 
     (\mu^{\psi}_j - \mu^{\psi}_l)^2
     \right] $ with $\sigma^{\psi}_{jl}$ denoting the covariance of the $j$-th and the $l$-th element of the $\psi$-transformed composition-valued mark.
     }
    \label{tab:testfunctions:CoDa}
\end{table}

Although in the following we restrict our discussion to only some characteristics, the same principle applies to all test functions in Table \ref{tab:testfunctions:CoDa}. 
Consider as an example  test function $\tf^{\psi,jl}_{1}$. It yields a componentwise  conditional mean product of marks $\tau_{jl}^{\psi}(r)$,
\begin{equation}\label{eq:componentwise:meanprodmarks:generic}
\tau_{jl}^{\psi}(r)=\mathds{E}_{\circ,r}\left[\psi_{j}(\cb(\circ))\psi_{l}(\cb(\mathbf{r}))\right],    
\end{equation}
which describes the average product of two components of the $\psi$-transformed marks for any pair of points as a function of the distance $r$. 

Restricting to pairs of the $j$-th element of $\psi(\cb)$ at two locations, $j=l$, and specifying $\psi$ in \eqref{eq:componentwise:meanprodmarks:generic} as log-ratio between parts $j_1$ and $j_2$, to directly focus on their relative contribution to the total,  yields the mean pairwise product of mark log-ratios $\tau_{jj}^{\mathrm{lr}}(r)$,
\begin{equation}\label{eq:logratio:meanprodmarks}
\tau_{jj}^{\mathrm{lr}}(r)=\mathds{E}_{\circ,r}\left[\log\left(\frac{c_{j_1}(\circ)}{c_{j_2}(\circ)}\right)\log\left(\frac{c_{j_1}(\mathbf{r})}{c_{j_2}(\mathbf{r})}\right)\right]
\end{equation}
for $j$ indexing the $\Tilde{D}=D^2$ ordered pairs $(j_1,j_2)=(1,1),\ldots,(1,D),\ldots,(D,D).$
This characteristic can highlight if the log-ratios at nearby points are correlated and their products thus tend to be larger (smaller) than expected under independence, $\nabla_{\tf_1}^{\mathrm{lr},jj}$.
It is easy to see that $\tau_{jj}^{\mathrm{lr}}$ equals the squared  mean $(\mu^{\mathrm{lr}}_{j})^2$ of the log-ratios of the $j_1$-th and $j_2$-th parts under independent marks.

Similarly, the componentwise log-ratio $\mathbf{r}$-mark functions $\tau^{\mathrm{lr}}_{j\bullet}$ and $\tau^{\mathrm{lr}}_{\bullet j}$ result from taking the conditional expectation of either $\tf^{\mathrm{lr}, jj}_{2}$ or $\tf^{\mathrm{lr}, jj}_{3}$, which both coincide with the mean of the log-ratio of the $j$-th part, $\mu^{\mathrm{lr}}_{j}$, in the case of independent marks. Reflecting the mean behaviour of the transformed mark at either the first or second point,  deviations of the empirical curves from $\mu^{\mathrm{lr}}_{j}$ at some distances indicate the presence of spatial structure in the mark parts, \ i.e. changes in the average mark ratios if a second point is present at distance $r$. As such, both quantities could be helpful tools to identify potential interrelations of the points and marks. 

Substituting $\tf^{\psi}_{4}$ for $\tf^{\psi}_{1}$ in \eqref{eq:componentwise:meanprodmarks:generic} yields a generic componentwise mark variogram \begin{equation}\label{eq:componentwise:vario:generic}
\gamma_{jl}^{\psi}(r)=\mathds{E}_{\circ,r}\left[0.5(\psi_{j}(\cb(\circ))-\psi_{l}(\cb(\mathbf{r})))^2\right].    
\end{equation}
The precise form again depends on the specific choice of $\psi$. Using a transformation into log-ratios and $j=l$ leads to a componentwise log-ratio mark variogram $\gamma_{jj}^{\mathrm{lr}}$ for composition-valued marks defined by
\begin{equation}
\label{eq:var_mark:lr}
\gamma_{jj}^{\mathrm{lr}}(r)=\mathds{E}_{\circ,r}\left[\frac{1}{2}\left(\log\left(\frac{c_{j_1}(\circ)}{c_{j_2}(\circ)}\right)-\log\left(\frac{c_{j_1}(\mathbf{r})}{c_{j_2}(\mathbf{r})}\right)\right)^2\right]
\end{equation}
for $j$ indexing $(j_1,j_2)=(1,1),\ldots,(1,D),\ldots,(D,D).$
Similar to the classic mark variogram, $\gamma_{jj}^{\mathrm{lr}}(r)$ concerns the average pairwise variation of the $j$-th log-ratio of $\psi(\cb)$ at two distinct points at distance $r$ and tends to the variance $\sigma_{jj}^{\mathrm{lr}}$ of the log-ratios of the $j_1$-st and $j_2$-nd parts for $r\to \infty$. The mark variogram can be used to investigate the heterogeneity among the marks, i.e.\ if the proportions of the specific parts for any pair of points are on average more similar in value for small distances.
We note that for each $r$ all of the above log-ratio characteristics could be stored in local $(D\times D)$ matrices with $D(D-1)$ log-ratios of different parts in the off-diagonal entries and zeros for the log-ratios of all parts with themselves on its diagonal. In particular, collecting all  $\gamma_{jj}^{\mathrm{lr}}(r)$ into a local mark variogram matrix $\boldsymbol{\Gamma}^{\mathrm{lr}}(r)$ is similar in spirit to a local variation matrix $\mathbf{T}(r)
$ 
which captures the spatial dispersion of the composition-valued mark at the distance $r$. 

While the above log-ratio characteristics are most useful for autocovariances and autocorrelations, the $\clr$ and $\ilr$ transformations allow for both auto- and cross-characteristic formulations. Choosing $\psi$ to denote the $\clr$ transformation, evaluation of the conditional expectation of $\tf^{\psi,jl}_{1}$ and $\tf^{\psi,jl}_{4}$ yields the conditional mean  product of clr marks,  $\tau^{\mathrm{clr}}_{jl}(r)$, and the clr mark variogram,  $\gamma^{\mathrm{clr}}_{jl}$, defined by 
\begin{eqnarray}\label{eq:clr:tau}
 \tau^{\mathrm{clr}}_{jl}(r)
 =
 \mathds{E}_{\circ,r}
 \left[\log\left(
 \frac{c_{j}(\circ)}{g(\cb)(\circ)}
 \right)
 \cdot
 \log\left(
 \frac{c_{l}(\mathbf{r})}{g(\cb)(\mathbf{r})}
 \right)
 \right]    
\end{eqnarray}
and
\begin{eqnarray}\label{eq:clr:vario}
\gamma^{\mathrm{clr}}_{jl}(r)
=
\mathds{E}_{\circ,r}
\left[
\frac{1}{2}
\left(\log
\left(
\frac{c_{j}(\circ)}{g(\cb)(\circ)}
\right)
-\log
\left(
\frac{c_{l}(\mathbf{r})}{g(\cb)(\mathbf{r})}
\right)
\right)^2
\right],
\end{eqnarray}
respectively. Due to the construction of \eqref{eq:clr:tau} and \eqref{eq:clr:vario}, both functions describe the average  spatial association/variation of the $j$-th and $l$-th parts relative to the geometric mean,
including both auto- (for $j=l$) and cross-relations (for $j\neq l$). In particular, cross-relations might be useful to analyse the association between different parts, e.g.\ the proportions of two distinct parasite species for neighbouring trees. Both quantities allow for similar interpretations as their log-ratio counterparts and could provide useful information on the distributional behaviour of the clr transformed parts. Inserting the corresponding terms into \eqref{eq:ExpectedIndMarksPsi} immediately implies that $\tau^{\mathrm{clr}}_{jl}(r)$ converges to the product of means $\mu^{\clr}_j\cdot\mu^{\clr}_l$ of the clr transformed parts $j,l$ for $r\rightarrow \infty$ as
\begin{align*}
    \nabla_{\tf_1}^{\clr,jl}
    &=
    \int_{\R}
    \int_{\R}
    \clr_j(\cb(\circ))\clr_l(\cb(\mathbf{r}))\varpi(\de \clr_j(\cb(\circ))\varpi(\de \clr_j(\cb(\mathbf{r})))\\
    &=
    \int_{\R}
    \clr_j(\cb)\varpi(\de \clr_j(\cb))\int_{\R}\clr_l(\cb)\varpi(\de \clr_l(\cb))\\
    &= \mu_j^{\clr}\mu_l^{\clr}
\end{align*}
Likewise, for $\gamma^{\mathrm{clr}}_{jl}(r)$, inserting $\tf_4^{\mathrm{clr}, jl}$ into \eqref{eq:ExpectedIndMarksPsi} yields 
\begin{align*}
    \nabla_{\tf_4}^{\clr,jl}
    &=  
    \int_{\R}
    \int_{\R}
    0.5(\clr_j(\cb(\circ))-\clr_l(\cb(\mathbf{r})))^2 
    \varpi(\de \clr_j(\cb(\circ))) \varpi(\de \clr_l(\cb(\mathbf{r})))
    \\
    &=
    0.5
    \Bigg[
    \int_{\R}
    (\clr_j(\cb))^2 \varpi(\de \clr_j(\cb))
    +
    \int_{\R}
    (\clr_l(\cb))^2 \varpi(\de \clr_l(\cb))
    \\
    &- 
    2\int_{\R}
    \int_{\R}
     \clr_j(\cb(\circ))\clr_l(\cb(\mathbf{r})) 
    \varpi(\de \clr_j(\cb(\circ)))
    \varpi(\de \clr_l(\cb(\mathbf{r})))
    \Bigg]
    \\
    &= 0.5\left[ 
    \sigma^{\mathrm{clr}}_{jj} + (\mu^{\mathrm{clr}}_j)^2 +
     \sigma^{\mathrm{clr}}_{ll} + (\mu^{\mathrm{clr}}_l)^2
     - 2 \mu^{\mathrm{clr}}_j \mu^{\mathrm{clr}}_l
     \right]
     \\
    &= 0.5\left[ 
    \sigma^{\mathrm{clr}}_{jj} + 
     \sigma^{\mathrm{clr}}_{ll} + 
     (\mu^{\mathrm{clr}}_j - \mu^{\mathrm{clr}}_l)^2 
     \right]
     \\
    &=:\zeta^{\mathrm{clr}}_{jl}. 
\end{align*}

The $\clr$ characteristics for all $D$ parts can efficiently be  stored in local $(D\times D)$ matrices including the local $\clr$ mark variogram matrix $\boldsymbol{\Gamma}^{\mathrm{clr}}(r)=\left[\gamma_{jl}^{\mathrm{clr}}(r)\right]_{j,l=1,\ldots,D}$.  
Similarly, using $\ilr$ coordinates yields
\begin{eqnarray}\label{eq:ilr:tau}
 \tau^{\mathrm{ilr}}_{jl}(r)
 =
 \mathds{E}_{\circ,r}
 \left[\ilr_j\left(\cb(\circ)\right)\ilr_l\left(\cb(\mathbf{r})\right)
 \right]    
\end{eqnarray}
and
\begin{eqnarray}\label{eq:ilr:vario}
\gamma^{\mathrm{ilr}}_{jl}(r)
=
\mathds{E}_{\circ,r}
\left[
\frac{1}{2}
\Bigg(\ilr_j(\cb(\circ))-\ilr_l(\cb(\mathbf{r}))\Bigg)^2
\right] ,
\end{eqnarray}
where $\ilr_j$ is the $j$-th ilr coordinate of the composition-valued marks and $\ilr_l$ analogous. 
\subsection{Compositional summary characteristics for composition-valued marks}\label{sec:CodaGlobal}

Instead of describing the spatial behaviour of one component of $\psi$-transformed compositions using a componentwise test function specification as covered in Table \ref{tab:testfunctions:CoDa}, we now discuss an extension which allows to evaluate the spatial properties of whole $D$-part compositions. 
This allows to assess the variation and correlation of the complete composition of e.g.\ nearby trees or business sectors. Such compositional summary characteristics, which summarise the variation and interrelation between the $D$-part compositions for a pair of points as a function of the interpoint distance $r$, can be derived using central concepts from the  Aitchison geometry. In particular, defining $\nabla_{\tf_f}^{\si}(r)$ to denote the conditional mean of the test function $\tf_f^{\si}$ of a $D$-part composition-valued mark $\cb$ at locations $\circ$ and $\mathbf{r}$ where $\Vert\circ-\mathbf{r}\Vert=r$, and   
$\kappa_{\tf_f}^{\si}(r)$ as
\begin{equation}\label{eq:ktf:fct:simplex}
 \kappa_{\tf_f}^{\si}(r)=\frac{\nabla_{\tf_f}^{\si}(r)}{\nabla_{\tf_f}^{\si}},   
\end{equation}
where $\nabla^{\si}_{\tf_f}$ extends \eqref{eq:ExpectedIndMarksPsi} to
\begin{eqnarray}\label{eq:ExpectedIndMarksCoda}
    \nabla_{\tf_f}^{\si}=
    \int_{\si^D}\int_{\si^D}\tf_f^{\si}(\cb(\circ),\cb(\mathbf{r})) \varpi(\de\cb(\circ)) \varpi(\de\cb(\mathbf{r}))
\end{eqnarray}
and denotes the conditional expectation of the test function for $r\to \infty$, allows to define different compositional mark summary characteristics. An overview of different test functions $\tf_f^{\si}$ and the corresponding characteristics is provided in Table \ref{tab:testfunctions:CoDa:global} where here only those test functions are considered which allow a one-to-one relation to the corresponding componentwise test functions of Table \ref{tab:testfunctions:CoDa}. We note that instead of applying \eqref{eq:ExpectedIndMarksPsi} to compute $\nabla_{\tf_f}^{\si}$ for the compositional versions of Schlather's and Shimantani's $I$ functions, $\nabla_{\tf_f}^{\si}$ is set to $\sigma^2_{\cb}$, 
\[
\sigma_{\cb}^{2} =\omega\sum^{\Tilde{D}}_{j=1} \zeta^{\psi}_{jj}
\]
where $\omega = 1/2D$ if a transformation into logratios is applied  and $\omega = 1$ under $\ilr$ and $\clr$ transformations to allow for a close analogy to Moran's $I$.  Like the componentwise mark characteristics, all of these test functions can be used to highlight particular aspects of the distributional properties.
However, providing different insights into the underlying structure of the composition-valued marks, the componentwise and the compositional mark characteristics both provide useful tools to investigate the individual contribution of the components to the results and to assess the overall variation and correlation.
\begin{table}
    \centering
    \begin{tabular}{lllll}
    \hline
     Name   & Test  & Normalising & Notation & Notation \\
    for $\tf_f^{\si}$   & function $\tf_f^{\si}$ &    factor $\nabla_{\tf_f}^{\si}$ & for $\nabla_{\tf_f}^{\si}(r)$ & for $\kappa^{\si}_{\tf_f}(r)$\\
    \hline
$\tf^{\si}_{1}$ & $\langle\cb(\circ), \cb(\mathbf{r})\rangle_A$ & $\mu_{\cb}^{2}$ & $\tau_{\cb\cb}(r)$ & $\kappa_{\cb\cb}(r)$ \\
  $\tf^{\si}_{4}$ & $0.5\Vert\cb(\circ)\ominus\cb(\mathbf{r})\Vert_A^2$ & $\sigma^{2}_{\cb}$ & $\gamma_{\cb\cb}(r)$ & $\gamma^{\mathrm{n}}_{\cb\cb}(r)$\\
    $\tf^{\si}_{5}$ & $\langle\cb(\circ)\ominus\cen(\cb),\cb(\mathbf{r})\ominus\cen(\cb)\rangle_A$ & $\sigma^{2}_{\cb}$ &  $\iota^{\mathrm{Shi}}_{\cb\cb}(r)$ &  $I^{ \mathrm{Shi}}_{\cb\cb}(r)$\\
    $\tf^{\si}_{6}$ & $\langle\cb(\circ)\ominus\cen(\cb)(\mathbf{r}),\cb(\mathbf{r})\ominus\cen(\cb)(\mathbf{r})\rangle_A$ & $\sigma^{2}_{\cb}$&  $\iota^{\mathrm{Sch}}_{\cb\cb}(r)$ &  $I^{ \mathrm{Sch}}_{\cb\cb}(r)$\\
   \hline
    \end{tabular}
    \caption{Compositional test function specifications with $\mu_{\cb}^{2} =\omega\sum^{\Tilde{D}}_{j=1} \mu^{\psi}_{j}\cdot\mu^{\psi}_{j}$ and $\sigma_{\cb}^{2} =\omega\sum^{\Tilde{D}}_{j=1} \sigma^{\psi}_{jj}$ with  $\omega = 1/2D$ if a transformation into logratios is applied and and $\omega = 1$ under $\ilr$ and $\clr$ transformations, $\cen(\cb)$ is the center of \eqref{eq:centerCoDa}, and $\cen(\cb)(\mathbf{r})$ is the conditional center computed over all pairs of points at a distance $\Vert \mathbf{r}\Vert=r$.
    }
    \label{tab:testfunctions:CoDa:global}
\end{table}
Noting the isometry of the Aitchison norm and distance   to their Euclidean counterpart versions of the clr or ilr transformed compositions, the test functions of Table \ref{tab:testfunctions:CoDa} and Table \ref{tab:testfunctions:CoDa:global} are related as follows. 
The compositional conditional expectation of the inner product of marks $\tau_{\mathbf{cc}}(r)$ can be constructed by using $\tf_1^{\si}$, which is related to the componentwise test function $\tf_1^{\psi, jl}$ through
\begin{eqnarray}\label{eq:equivtf1}
\tf_1^{\si} & = \langle \cb(\circ), \cb(\mathbf{r}) \rangle_A =&  \langle \clr(\cb(\circ)), \clr(\cb(\mathbf{r})) \rangle_E\\\nonumber
    & = \sum^{D}_{j=1} \clr_j(\cb(\circ)) \clr_j(\cb(\mathbf{r})) =& \sum^{D}_{j=1} \tf_1^{\clr, jj}  \\ \nonumber
    & = \sum^{D-1}_{j=1} \ilr{_j}(\cb(\circ))\ilr{_j}(\cb(\mathbf{r})) =& \sum^{D-1}_{j=1} \tf_1^{\ilr, jj}\\ \nonumber
     \overset{\eqref{eq:AitInner}}{=}& \frac{1}{2D}\sum_{j_1}\sum_{j_2} \log\left(\frac{c_{j_1}(\circ)}{c_{j_2}(\circ)}\right)\log\left(\frac{c_{j_1}(\mathbf{r})}{c_{j_2}(\mathbf{r})}\right)  =& \frac{1}{2D}\sum^{D^2}_{j=1} \tf_1^{\mathrm{lr}, jj}
\end{eqnarray}
with $\langle\cdot,\cdot\rangle_E$ denoting the Euclidean inner product. Thus 
\begin{eqnarray}\label{eq:relation:tau:global:component}
\tau_{\mathbf{cc}}(r)=\sum^{D}_{j=1} \tau_{jj}^{\clr}(r)=\sum^{D-1}_{j=1}\tau_{jj}^{ilr}(r)=\frac{1}{2D}\sum^{D^2}_{j=1}\tau_{jj}^{\mathrm{lr}}(r)
\end{eqnarray}
which allows to evaluate which individual components contribute to the overall mark characteristic. 
The corresponding compositional mark correlation function $\kappa_{\mathbf{cc}}$ follows by normalising $\tau_{\mathbf{cc}}$ by $\mu_{\cb}^{2}$, 
\begin{eqnarray}\label{eq:imit:muc}
     \mu_{\cb}^{2} &=& \int_{\si^D}\int_{\si^D}\langle \cb(\circ), \cb(\mathbf{r}) \rangle_A \varpi(\de\cb(\circ)) \varpi(\de\cb(\mathbf{r}))\\ \nonumber
   & = & \omega \int_{\R^{\Tilde{D}}}\int_{\R^{\Tilde{D}}} \langle \psi(\cb(\circ)), \psi(\cb(\mathbf{r})) \rangle_E \varpi(\de\psi(\cb(\circ))) \varpi(\de\psi(\cb(\mathbf{r})))  \\ \nonumber
   & =& \omega\sum^{\Tilde{D}}_{j=1}\int_{\R^{\Tilde{D}}}\int_{\R^{\Tilde{D}}} \psi_j(\cb(\circ)), \psi_j(\cb(\mathbf{r})) \varpi(\de \psi_j(\cb(\circ))) \varpi(\de\psi_j(\cb(\mathbf{r})))  \\ \nonumber
   & =& \omega\sum^{\Tilde{D}}_{j=1} \mu_j^{\psi}\cdot\mu_j^{\psi} \nonumber,
\end{eqnarray}
the limiting case which follows by substituting $\tf_1^{\si}$ for $\tf_f^{\si}$ in \eqref{eq:ExpectedIndMarksCoda} with $\omega = 1/2D$ under the lr-transformation and $\omega = 1$ if a transformation into $\ilr$ or $\clr$ coordinates is applied. Hence, for independent composition-valued marks, $\kappa_{\mathbf{cc}}$ becomes constant one. 

Selection of $\tf_4^{\si}$ results in a compositional  mark variogram,  $\gamma_{\mathbf{cc}}(r)$,
\begin{equation}\label{eq:coda:gamma}
   \gamma_{\mathbf{cc}}(r) = \e_{\circ,r}\left[\frac{1}{2}\|\cb(\circ)\ominus\cb(\mathbf{r})\|_A^2\right] = \e_{\circ,r}\left[\frac{1}{2}d_A(\cb(\circ),\cb(\mathbf{r}))^2\right].
\end{equation}
By the decomposition of the Aitchison squared  distance into the sum of the squared distances between the $\clr$ or $\ilr$ transformed marks as stated in \eqref{eq:AitDist}, the compositional mark variogram can be decomposed into the sum of the componentwise mark variogram terms. In particular, we have that 
\[
\gamma_{\mathbf{\cb\cb}}(r)=\sum^D_{j=1}\gamma_{jj}^{\mathrm{clr}}(r)=\sum^{D-1}_{j=1}\gamma_{jj}^{\mathrm{ilr}}(r).
\]
This functional quantity describes the average dispersion between the $D$-part composition-valued marks for a focal and a second point at distance $r$. 
If composition-valued marks are correlated for small distances, the compositional mark variogram $\gamma_{\mathbf{cc}}(r)$ will show a corresponding decrease for small spatial distances. The decomposition allows to evaluate which mark components contribute how to the overall compositional mark variogram $\gamma_{\mathbf{cc}}(r)$.   
Again using  \eqref{eq:ExpectedIndMarksCoda}, the mark variogram under independent marks is equal to $\sigma^{\si}_{\cb}$,
\begin{eqnarray}
\sigma^{\si}_{\cb} & = & 0.5\int_{\si^D}\int_{\si^D} \Vert \cb(\circ)\ominus\cb(\mathbf{r}) \Vert^2_A \varpi(\de\cb(\circ)) \varpi(\de\cb(\mathbf{r})) \\ \nonumber
   & = &  0.5\omega \int_{\R^{\Tilde{D}}}\int_{\R^{\Tilde{D}}} \Vert \psi(\cb(\circ))\ominus\psi(\cb(\mathbf{r})) \Vert_E^2 \varpi(\de\psi(\cb(\circ))) \varpi(\de\psi(\cb(\mathbf{r})))   \\ \nonumber
   & = &0.5 \omega\sum^{\Tilde{D}}_{j=1}\int_{\R^{\Tilde{D}}}\int_{\R^{\Tilde{D}}}  \left(\psi_j(\cb(\circ))-\psi_j(\cb(\mathbf{r}))\right)^2  \varpi(\de\psi_j(\cb(\circ))) \varpi(\de\psi_j(\cb(\mathbf{r})))  \\ \nonumber
   & = & \omega \sum^{\Tilde{D}}_{j=1} \zeta^{\psi}_{jj}\nonumber.
\end{eqnarray}
 
Test functions $\tf_5^{\si}$ and $\tf_6^{\si}$ yield adaptations of Schlather's and Shimatani's $I$ functions (denoted by $\iota_{\cb\cb}^{\mathrm{Sch}}$ and $I_{\cb\cb}^{\mathrm{Sch}}$ and $\iota_{\cb\cb}^{\mathrm{Shi}}$ and $I_{\mathbf{\cb\cb}}^{\mathrm{Shi}}$), respectively, which can be useful tools to investigate the spatial autocorrelation of the composition-valued marks. 
Writing $\Bar{\cb}=\cb-\cen(\cb)$ to denote the centered composition,  $\iota_{\cb\cb}^{\mathrm{Shi}}$ can be decomposed into the weighted sum over the componentwise Shimantani's functions $\iota_{\cb\cb}^{\psi,jl}$ analogously to \ref{eq:relation:tau:global:component}, using the equivalences of the Aitchinson inner product of  
\eqref{eq:equivtf1}. The unnormalised $\iota_{\cb\cb}^{\mathrm{Sch}}$ function of Schlather can be obtained from its componentwise versions in a similar manner by applying a centering of the composition by the conditional center $\cen(\cb)(\mathbf{r})$.
The decomposition of $\iota_{\cb\cb}^{\mathrm{Sch}}(r)$ 
into individual contributions is also analogous to that for $\gamma_{\cb\cb}(r)$. 

\subsection{Extensions to composition-valued marks with total information}\label{sec:CodaTotal}

Finally, extensions of the proposed mark characteristics to combinations of relative, i.e.\ composition-valued, and absolute point-specific information are outlined. Adapting the results of  \cite{codatotal} to the present context, consider $\tilde{\cb}\in \R^D$ such that $\cb=\cls(\tilde{\cb})\in \si^D$ and denote by $y=\sum_{j=1}^{D}\tilde{c}_j$ the total. We then call $\boldsymbol{\eta}=(y,\cb)$ a mixed mark with real-valued and composition-valued components $y$ and $\cb$ living on $\mathds{T}=\mathds{R}_+\times \mathds{S}^D,~\R_+ = \left[0,\infty\right)$. Focusing on the process $\lbrace x_i, \boldsymbol{\eta}(x_i)\rbrace$ instead of $\lbrace x_i, \cb(x_i)\rbrace$, all the above mark characteristics can be extended to mixed 
marks on $\mathds{T}$. To this end, consider first the scalars $y,y^{\prime}\in\R_+$ and let $\oplus_+$ and $\odot_+$ denote the plus-perturbation and plus-powering operations, respectively, where 
$y\oplus_+ y^{\prime}=yy^{\prime}$ and $\xi\odot_+ y=y^{\xi}$ for $\xi\in\R$. Further, denote by $\langle y,y^{\prime}\rangle_+=\langle\log(y),\log(y^{\prime})\rangle_E$ the plus-inner product and  by $d_+(y,y^{\prime})=\mathrm{abs}(\log\left(y\right)-\log\left(y^{\prime}\right))$ the plus-distance on $\R_+$. The above results can then be used to establish a vector space structure on $\mathds{T}$ with $\mathds{T}$-perturbation  $\boldsymbol{\eta}\oplust \boldsymbol{\eta}=(y\oplus_+ y^{\prime},  \cb\oplus \cb^{\prime})$ and  $\mathds{T}$-powering operations $\xi\odott\boldsymbol{\eta}^{\prime}=(y^\xi, \xi\odot\cb)$ where $\boldsymbol{\eta},\boldsymbol{\eta}^\prime\in\mathds{T}$ and $\xi\in\R$ as before. Both variation and correlation related mark characteristics from Section  \ref{sec:CodaGlobal} can be redefined through the $\mathds{T}$-inner product 
and $\mathds{T}$-squared distance 
\begin{eqnarray*}
\langle\boldsymbol{\eta},\boldsymbol{\eta}^{\prime}\rt
=
\langle\cb,\cb^{\prime}\rangle_A 
+
\beta\langle y,y^{\prime} \rangle_+ 
\end{eqnarray*}
and  
\begin{eqnarray*}
    d_{\mathds{T}}(\boldsymbol{\eta},\boldsymbol{\eta}^{\prime})
    =
    d_A(\cb,\cb^{\prime})
    +
    \beta\left(
    \mathrm{abs}(
    \log
    \left(y\right)
    -
    \log
    \left({y^{\prime}}\right)
    \right)),
\end{eqnarray*}
respectively, where $\beta$ is a weight which can e.g. be chosen as one or as the ratio of the variances of the composition $\cb$ and of the total $y$ \citep{HappGrevenJASA}. Using the above extensions and reformulating the compositional  mark variogram for mixed composition- and real-valued marks, we obtain
\begin{eqnarray}\label{eq:mixed:vaio}
  \gamma_{\cb\cb,y}(r) = 
  \e_{\circ,r}\left[
   \frac{1}{2}d_A(\cb(\circ),\cb(\mathbf{r}))^2
 +\beta(\frac{1}{2}d_+(y(\circ),y(\mathbf{r})))^2.
  \right]
\end{eqnarray}
The other summary characteristics of Table \ref{tab:testfunctions:CoDa:global} can be extended analogously.

\subsection{Estimation}\label{sec:estimation}

Recalling the representation of $\nabla^{\psi, jl}_{\tf_f}(r)$ as the ratio of the two second-order product density functions $\varrho^{\psi,(2)}_{\tf_f}(r)$ and $\varrho^{(2)}(r)$,
$\nabla^{\psi, jl}_{\tf_f}(r)$ can be estimated in close analogy to classic spatial point processes by  
\begin{equation}\label{eq:nabla:est}
  \widehat{\nabla^{\psi,jl}_{\tf_f}}(r)=\widehat{\varrho^{\psi,jl,(2)}_{\tf_f}(r)}\left/\widehat{\varrho^{(2)}_{\tf_f}(r)}\right.  
\end{equation}
where 
\begin{equation}\label{eq:est:varrho:marked}
\widehat{\varrho^{\psi,jl,(2)}_{\tf_f}}(r)=\frac{1}{2\pi r \nu(W)}\sum^{\neq}_{x_1, x_2\in W}
\tf^{\psi,jl}_f(\psi_j(\mathbf{c}(x_1)), \psi_l(\mathbf{c}(x_2)))\mathfrak{K}_b(\Vert x_1-x_2\Vert-r) 
\end{equation}
and
\begin{equation}\label{eq:est:varrho}
 \widehat{\varrho^{(2)}}(r)=\frac{1}{2\pi r \nu(W)}\sum^{\neq}_{x_1, x_2\in W}
\mathfrak{K}_b(\Vert x_1-x_2\Vert-r) .   
\end{equation}
Here, $\mathfrak{K}_b$ denotes a kernel function of bandwidth $b$ and $\nu(W)$ the area of the observation window $W$. We note that as \eqref{eq:est:varrho:marked} and \eqref{eq:est:varrho} are estimated using the same estimation principle, an edge correction factor can be ignored in both expressions  \citep{Illian2008}.
Similarly, an estimator of $\kappa^{\psi, jl}_{\tf_f}$ of \eqref{eq:ktf:fct:psi} can be obtained from normalizing $\widehat{\nabla^{\psi, jl}_{\tf_f}}(r)$ by $\widehat{\nabla^{\psi, jl}_{\tf_f}}$,      
\begin{equation}\label{eq:ktf:psi:estimator}
 \widehat{\kappa^{\psi, jl}}_{\tf_f}(r)=\widehat{\nabla^{\psi, jl}_{\tf_f}}(r)\left/\widehat{\nabla^{\psi, jl}_{\tf_f}}\right.,   
\end{equation}
where $\widehat{\nabla^{\psi, jl}_{\tf_f}}$ in  \eqref{eq:ktf:psi:estimator} can be estimated analogously to the scalar case from the transformed marks of the $n$ points $x_1,\ldots, x_n$ by
\[
\widehat{\nabla^{\psi, jl}_{\tf_f}}=\frac{1}{n^2}\sum_{i=1}^n\sum^n_{h=1}\tf^{\psi, jl}_f(\psi_j(\mathbf{c}(x_i)), \psi_l(\mathbf{c}(x_h)))
\]
\citep{Illian2008}.
For example,  
specifying $\psi_j$ 
as the $j$-th component of the $\clr$ transformation of $\mathbf{c}$, i.e.\ $\log(c_j(\circ)/g(\mathbf{c}))$, an estimator of the clr mark variogram $\gamma^{\clr}_{jl}$ of \eqref{eq:clr:vario} can be obtained from the ratio of second-order density functions $\widehat{\varrho^{\clr, jl,(2)}_{\tf_4}(r)}/\widehat{\varrho^{(2)}_{\tf_4}(r)}$, 
\[
\widehat{\gamma^{\clr}_{jl}}=\frac{\sum^{\neq}_{x_1, x_2\in W}
0.5\left(\log\left(\frac{c_j}{g(\mathbf{c})}\right)(x_1)-\log\left(\frac{c_l}{g(\mathbf{c})}\right)(x_2)\right)^2\mathfrak{K}_b(\Vert x_1-x_2\Vert-r)}{\sum^{\neq}_{x_1, x_2\in W}\mathfrak{K}_b(\Vert x_1-x_2\Vert-r)},
\]
as constant terms cancel. Likewise, considering a transformation into logratios, an estimator of the mean pairwise product of mark log-ratios $\tau_{jj}^{\mathrm{lr}}(r)$ of \eqref{eq:logratio:meanprodmarks} is obtained as 
\[
\widehat{\tau^{\mathrm{lr}}_{jj}}=\frac{\sum^{\neq}_{x_1, x_2\in W}
\left(\log\left(\frac{c_{j_1}}{c_{j_2}}\right)(x_1)\cdot\log\left(\frac{c_{j_1}}{c_{j_2}}\right)(x_2)\right)\mathfrak{K}_b(\Vert x_1-x_2\Vert-r)}{\sum^{\neq}_{x_1, x_2\in W}\mathfrak{K}_b(\Vert x_1-x_2\Vert-r)}.
\]

Considering the unmarked case first and applying the Campbell theorem \citep{Chiu2013} we have
\begin{eqnarray*}
 \e\left[\widehat{\varrho^{(2)}}(r)\right]&=\int \mathfrak{K}_b(s)\varrho^{(2)}(r+bs)\de s.
\end{eqnarray*}
Noting that $ \e\left[\widehat{\varrho^{(2)}}(r)\right]\rightarrow\varrho^{(2)}$ as $b\rightarrow 0$ it follows that \eqref{eq:est:varrho} is an unbiased estimator for $b \rightarrow 0$ of the second-order product density function. Analogously, $\widehat{\varrho_{\tf_f}^{\psi,jl,(2)}}$ of \eqref{eq:est:varrho:marked} can be shown to be an unbiased estimator of $\varrho_{\tf_f}^{\psi,jl,(2)}$ for $b\rightarrow 0$ by applying the Campbell theorem to the marked case \citep{Daley2003} such that $ \e\left[\widehat{\varrho_{\tf_f}^{\psi,jl,(2)}}(r)\right]\rightarrow\varrho_{\tf_f}^{\psi,jl,(2)}$. As both \eqref{eq:est:varrho:marked} and \eqref{eq:est:varrho} yield unbiased estimators, \eqref{eq:nabla:est} yields a ratio-unbiased estimator for $b\rightarrow 0$.

\section{Test of random labeling hypothesis for composition-valued marked point processes}\label{sec:testing}

To test for deviations from the null hypothesis of random labels, i.e.\ marks that are $i.i.d.$ and thus independent of each other and the points, we adopt global envelope tests. These are  non-parametric tests based on $s$ simulations of the test statistic under the null model, originally introduced by \cite{mari1} to solve multiple testing problems in spatial statistics.  
In the case of the random labeling hypothesis, the simulations can be obtained simply by permuting the marks of the points \citep[e.g.][]{myllymaki_deviation_2015}. In the case of composition-valued marks, we take the same approach, i.e.\ permute the composition-valued marks. 

Thus, first, the marks are permuted $s$ times. The next step then is  to compute the test statistic  from the observed marked point pattern $\lbrace (x_i, \cb(x_i) )\rbrace^n_{i=1}$ and the $s$ simulated patterns with permuted composition-valued marks.
Let $\vartheta_1(r)$ stand for the empirical functional test statistic   computed from the observed pattern and let $\vartheta_2(r), \dots, \vartheta_{s+1}(r)$ be the $s$ functional test statistics computed from the $s$ simulated patterns. 
Then a Monte Carlo test is done based on $\vartheta_1(r), \dots, \vartheta_{s+1}(r)$. 
If the functional test statistics can be ordered from the least extreme to the most extreme, a Monte Carlo $p$-value can be computed for the test in a similar manner as in the classical Monte Carlo test \citep{barnard1963}.
Here, to order the statistics, we use the extreme rank length (ERL) measure \citep{mari1, MrkvickaEtal2020} as a particular instance of rank-based measures which also allow for the graphical interpretation of the test in terms of a global envelope. 
Please refer to the publications citet above and \citet[][Appendix A]{GETpack} for the definition of ERL and a discussion of alternative rank measures.

More precisely, the idea of the global envelope test is the following: Let us denote by $E_i,~i= 1,\ldots, s+1$, the measure associated with the $i$-th functional test statistic. Let $\prec$ be an ordering for the measures $E_i$ such that $E_i\prec E_j$ whenever $\vartheta_i$ is more extreme than $\vartheta_j$ with respect to the measure $E$. The critical value $E_{(\alpha)}$ under a given significance level $\alpha$ can then be found as the largest $E_i$ which satisfies
\[
\sum_{i=1}^{s+1} \mathds{1}(E_i\prec E_{(\alpha)})\leq \alpha(s+1).
\]
Let us then denote by $I_{(\alpha)}$ the index set of the 
test statistics $\vartheta_i$ that are less or as extreme as $E_{(\alpha)}$ as measured by their associated $E_i$. 
Then the $100(1-\alpha)$\% global envelope is the band given by the two 
functions
$$
\vartheta^{l}_{(\alpha)}(r)=\min_{i \in I_{(\alpha)}}\vartheta_{i}(r)
$$ 
and 
$$
\vartheta^{u}_{(\alpha)}(r)=\max_{i \in I_{(\alpha)}}\vartheta_{i}(r).
$$ 
If $\vartheta_1(r)$ goes outside of the envelope $(\vartheta^{l}_{(\alpha)}(r),\vartheta^{u}_{(\alpha)}(r))$ for any of its argument values $r$, there is evidence to reject the null hypothesis at the given significance level $\alpha$. Further, the values of $r$ where $\vartheta_1(r)$ leaves the envelope show the reasons of the rejection of the test. 
There is a one-to-one correspondence between the graphical interpretation by the global envelope and the Monte Carlo $p$-value of the test,  given by
\begin{eqnarray}\label{eq:p}
p = \frac{1}{s+1} \left\{ 1 + \sum_{i=2}^{s+1} \mathbf{1}(E_i\prec E_1) \right\}.
\end{eqnarray}
That is, assuming that there are no pointwise ties in $\vartheta_i(r)$, $i=1,\dots,s+1$, with probability 1, the empirical statistic $\vartheta_1(r)$ leaves the envelope if and only if $p\leq \alpha$, and $\vartheta^{l}_{(\alpha)}(r) \leq \vartheta_1(r)\leq \vartheta^{u}_{(\alpha)}(r)$ if $p > \alpha$ \citep[][Theorem 1]{mrkvickaEtal2021}.
The size of the test based on the $p$-value \eqref{eq:p}, and thus the global envelope, is $\alpha$ when the test statistics $\vartheta_i(r)$ can be strictly ordered and $\alpha(s+1)$ is an integer \citep[][Lemma 1]{mari1}.
In practice, the functional test statistics are estimators of functional summary characteristics computed on a chosen finite but dense set of argument values $r$.
If the random labeling hypothesis concerns all $D$ parts of the composition-valued marks, the functional test statistics could be specified by $\vartheta(r) = \widehat{\kappa_{\tf_f}^{\si}}(r)$ for any of the functional test statistics in Table \ref{tab:testfunctions:CoDa:global}.
Such a test can point out distances $r$ which are responsible for the potential rejection of the test, but
does not focus on which components in the composition-valued marks show the strongest dependence.

Alternatively, the functional test statistic can be constructed from the componentwise mark summary characteristics using the combining procedure of \citet[][Appendix B]{GETpack}. For example, for the componentwise mark variograms, i.e. $\gamma_{jj}^{\psi}(r)$, $j=1,\dots,\tilde{D}$, and a set of $d$ $r$-values,
the test vector is constructed from the estimated mark variograms $\widehat{\gamma_{jj}^{\psi}}(r)$, $j=1,\dots,\tilde{D}$:
\begin{eqnarray}\label{eq:test_combined}
\boldsymbol{\vartheta} = \left((\widehat{\gamma^{\psi}}_{11}(r_1), \dots, \widehat{\gamma^{\psi}}_{11}(r_d)), \dots, (\widehat{\gamma^{\psi}}_{\tilde{D}\tilde{D}}(r_1), \dots,  \widehat{\gamma^{\psi}}_{\tilde{D}\tilde{D}}(r_d))\right).
\end{eqnarray}
This test summarizes the information from all the components of the composition-valued marks using the chosen componentwise characteristics. This test holds the global significance level for the complete test vector \eqref{eq:test_combined} and it can point out both the components $j$ and distances $r$ which are responsible for the potential rejection of the test.


\section{Applications}\label{sec:appl}

We use our new mark characteristics  to analyse  two marked point patterns from forestry and urban economics. In particular, focusing on our extensions of three prominent characteristics from the literature with each of these addressing a different aspect of the marks, we utilized the mark variogram, the conditional mean product of the marks and Shimantani's $I$ function to investigate the variation, association and autocorrelation of the composition-valued marks.  
The forestry data is an example of a spatial point pattern with $2$-part composition-valued marks and additional absolute information (totals), while the urban economics data includes compositional marks with 4 parts.
%

\subsection{Application to tree data with crown-to-base proportion}

\begin{figure}[h!]
    \centering
    \includegraphics[scale=.5]{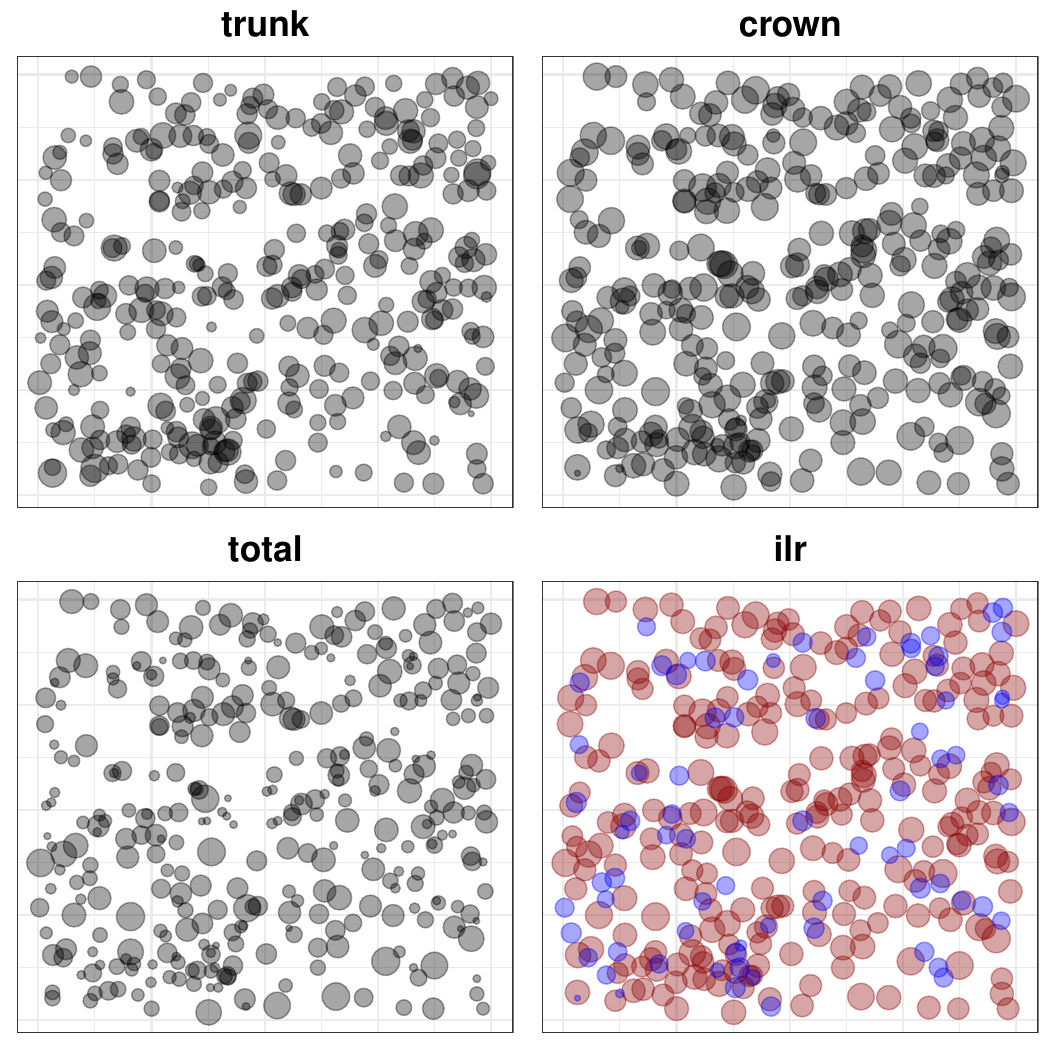}
    \caption{Spatial distribution of the marks and ilr-transformed composition-valued marks  for the Finnish forest stand data. Base (top left) and crown (top right) proportions and absolute height in metres (bottom left)  per tree with the diameter of the discs proportional to the mark values. The ilr-transformed crown-to-base proportions (bottom right) with negative values shown in blue and positive values  highlighted in red.}
    \label{fig:treepat}
\end{figure}

The data on forest tree stands under study originates from a forest development study of managed, uneven-aged Norway spruce forests conducted under the ERIKA research project at the Natural Resources Institute Finland (Luke) \citep{Eerik2007,Eerik2014,SAKSA20111409}. 
Several plots of size $40\text{ m} \times 40\text{ m}$ were recorded in southern Finland. Here we analyse a plot with 349 trees located in Vesijako (see Figure \ref{fig:treepat}).
The tree locations and associated tree characteristics, including the total height of the tree, $h$, and the height from the ground to the first living branch (of the crown), $h_{b}$, were recorded for all trees with $h\geq 1.3$ m, which corresponds to the height at which the diameter at breast height (DBH) of a tree is measured. The height of the crown was obtained by $h_{cr} = h-h_{b}$. In what follows, we call $h_{b}$ 'base' for short.
Instead of the absolute values of $h_{cr}$ and $h_b$ 
(which clearly depend on the individual age of the trees), 
we considered the relative base height $h_{b}/h$ (with geometric mean 0.37), relative crown height $h_{cr}/h$  (with geometric mean 0.58) and the total height $h$.
A summary of these three marks (two relational, one absolute) is provided in Table \ref{tab:summaryTree}. 
\begin{table}[]
    \centering
        \begin{tabular}{lrrrrrr}
\hline
    Mark & Min & 1st quantile & Mean & Median & 3rd quantile & Max\\
        \hline
$h_b/h$ & 0.11 & 0.30 & 0.39 & 0.40 & 0.48 & 0.81\\ 
$h_{cr}/h$ & 0.19 & 0.52 & 0.61 & 0.60 & 0.70 & 0.89 \\ 
\hline
$h$ & 1.36 & 3.19 & 6.30 & 8.10 & 12.50 & 26.2\\  
  \hline
  \end{tabular}
    \caption{Summary statistics for the relative height from the ground to the first living branch of the tree ($h_{b}/h$), the relative height of the crown ($h_{cr}/h$), and the total height ($h$) in meters for the Finnish tree data.\label{tab:summaryTree}}
\end{table}
The total tree heights varied up to 26.2 meters with an average crown proportion of 61\%.
The spatial distribution of the marks and the corresponding $\ilr$ coordinates are depicted in Figure \ref{fig:treepat}.
Crown proportions tended to be rather large, with only some small values (see top right panel of Figure \ref{fig:treepat}).
Compared to the crown-to-base composition, 
there is greater variation in the tree heights (bottom left). 
The $\ilr$ coordinates (bottom right) corresponding to the log-ratio transformation of the crown-to-base compositions support the above impressions: Most of the $\ilr$ coordinates are positive (indicated in red), which occurs when crown proportions are larger than base proportions. 

Initially we considered only the crown-to-base ratios through their $\ilr$ coordinates and conducted a separate analysis where the total height was examined at its original scale.
Next, to extend the composition-valued mark by the absolute height information, we additionally included the log transformation of the total heights in a vector-valued mark in our computations according to Section \ref{sec:CodaTotal}. For both steps of our analysis and each mark characteristic, we computed the 95\% global envelope under the random labeling hypothesis (see Section \ref{sec:testing} for details), based on 3000 permutations. While the interplay of tree crowns and total heights, and competition between neighbouring trees have been of interest in different studies \citep{Hegyi74, https://doi.org/10.1002/bimj.4710350412, HUI201849, PITKANEN2022102941}, we are not aware of studies on the interdependencies of crown-to-base proportions and, additionally, the total heights over space. In particular, different from the existing approaches, the proposed rescaling of the absolute information into relative proportions allows for the characterisation of the structural properties of the marks  without being affected by any heterogeneity in the age or types of the trees under study.

\begin{figure}[h!]
    \centering
      \includegraphics[scale=.65]{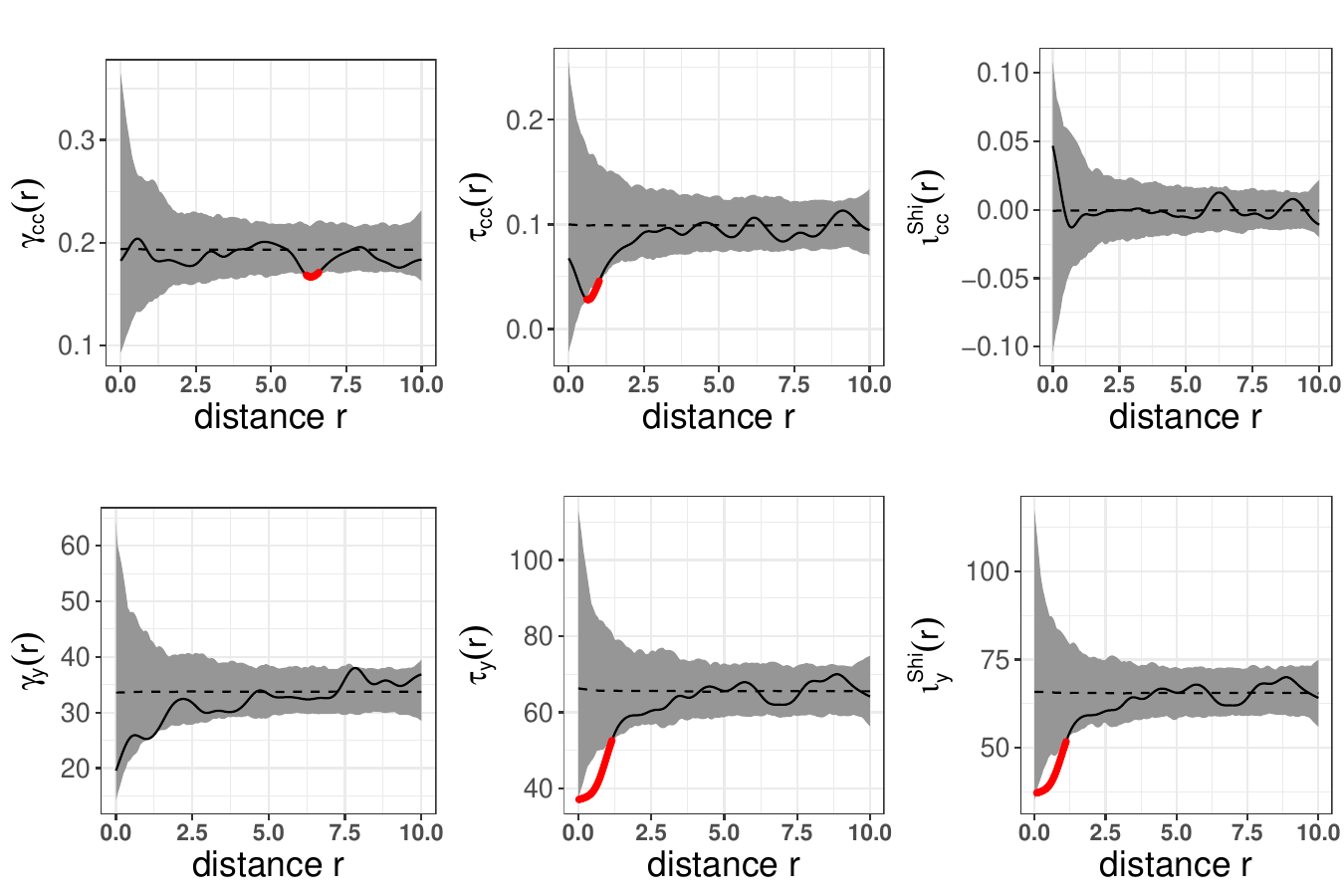}
   \caption{Selected compositional mark summary characteristics and 95\% global envelopes (shaded) based on 3000 permutations of marks computed from the ilr transformed crown-to-base proportions (top row) and the total height (bottom row): mark variogram $\gamma_{\cb\cb}$ (top left), the conditional mean product of marks $\tau_{\cb\cb}$ (top central), Shimantani's $\iota_{\cb\cb}^{\mathrm{Shi}}$ (top right), mark variogram $\gamma_{y}$ (bottom left), the conditional mean scalar product of marks $\tau_{y}$ (bottom central) and Shimantani's $\iota_{y}^{\mathrm{Shi}}$ (bottom right). The dashed line corresponds to the mean function under the random labeling hypothesis  and the solid curve to the test function on the observed data. Distances are given in meters.}
\label{fig:results:tree}
\end{figure}

The top row of Figure \ref{fig:results:tree} shows the compositional mark variogram $\gamma_{\cb\cb}$ (left), the conditional mean product of marks $\tau_{\cb\cb}$ (central) and Shimantani's $\iota_{\cb\cb}$ (right)  together with 95\% global envelopes. We note that for $D=2$ as in this application, all three compositional characteristics coincide with their componentwise analogues using the ilr transformation. 
While the first two empirical characteristics for the crown-to-base composition leave the global envelope for some distances $r$, the third one
is completely inside the global envelope. The mark variogram suggests that the average dispersion of the crown-to-base log-ratios is smaller than expected under the random labeling hypothesis for any pair of points at interpoint distances of about $r=6$ m. This would correspond to above random similarity in the crown-to-base log-ratios for neighbouring trees at intermediate distances.
The results for $\tau_{\cb\cb}$ suggest that 
the product of the crown-to-base log-ratios of two points at distance $r\approx 1$ m apart from each other 
tend to be smaller than expected under the random labeling hypothesis. 
Recall that small values of $\tau_{\cb\cb}(r)$ occur for distance $r$ if the transformed tree compositions for any two points at a distance $r$ are more different, e.g.\ trees with large crown proportions (i.e.\ positive coordinates) are surrounded by trees with small crown proportions (i.e.\ negative coordinates). This finding might be explained by crown competition and growth restrictions due to space limitations for closely neighbouring trees.
Reinspecting the $\ilr$ scores of Figure \ref{fig:treepat}, positive values (red, i.e. large crown to small base ratio) occur in close distance to negative $\ilr$ coordinates (blue, i.e.\ small crown to large base ratio) which could explain the results.  

Comparing the findings with the results for the total information depicted in the bottom row of Figure \ref{fig:results:tree}, both the conditional mean product of marks $\tau_{y}$ and Shimantani's $\iota_{y}^{\mathrm{Shi}}$ show clear negative deviations from the global envelopes for distances $r < 1.25$ m. These findings indicate that on average large trees are surrounded by smaller trees at shorter distances and vice versa implying negative autocorrelation potentially due to competition. On the other hand, the mark variogram $\gamma_{y}$ is completely within the 95\% global envelope, even though it is rather close to the lower boundary for small $r$. Thus, no significance dependence was detected using the squared difference of the tree heights as the test function.


Additionally, specifying the weight $\beta$ introduced in Section \ref{sec:CodaTotal} as the ratio of the variances for the ilr-transformed composition and the log-transformed totals, we computed all three mark characteristics following the concepts  outlined in Section \ref{sec:CodaTotal}. Accounting for the total information in the analysis of the crown-to-base composition, all three characteristics are completely covered within the global envelopes. This result means that when taking both mark components $\cb$ and $y$ jointly into account, the marks do not show any significant spatial dependence or autocorrelation. 
 \begin{figure}[h!]
    \centering
      \includegraphics[scale=.65]{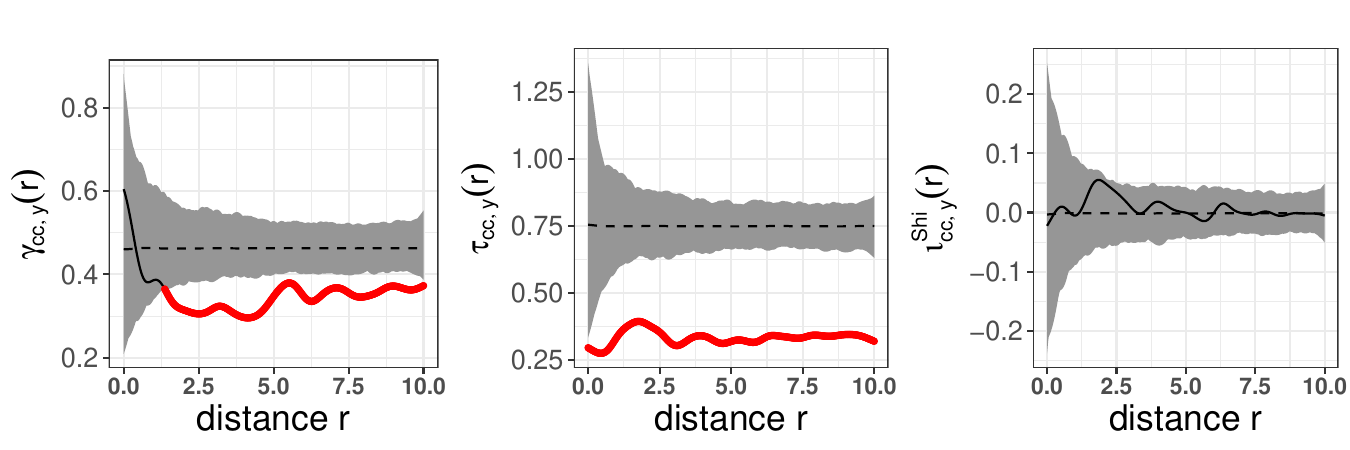}
   \caption{Selected compositional mark summary characteristics and envelopes (shaded) based on 3000 simulations computed from the ilr transformed crown-to-base proportions and log transformed total height as mixed mark with $\beta = 0.57$. Mark variogram $\gamma_{\cb\cb, y}$ (left), the conditional mean inner product of marks $\tau_{\cb\cb, y}$ (central) and Shimantani's $\iota_{\cb\cb, y}^{\mathrm{Shi}}$ (right). The dashed line corresponds to the mean function under the random labeling hypothesis and the solid curve to the test function on the observed data. Distances are given in meters. }
\label{fig:results:tree:total}
\end{figure}

\subsection{Application to Spanish municipalities data with local business sector compositions}

As a second application, we considered data from the National Statistics Institute of Spain (INE) on a  de-composition of the local economy into four different sectors at municipality level. The data at hand was generated using a data query at the official webpage (\url{www.ine.es}) and derives directly from information collected in the Spanish central business register. The decomposition into the four distinct sectors was available for Spanish municipalities with at least 1,000 inhabitants and refers to the total number of all economic actors, e.g.\  companies, with location in a given municipality. To protect against  inconsistencies in the data assignment and potential problems with multiple spatially wide spreading business locations,  each company was matched by INE in a pre-processing step to exactly one municipality using the address information of the corporate headquarter.
In subsequent steps, each company was categorised into  one of the four business sectors
\begin{enumerate}
    \item[(a)] \textbf{industry} (including extractive and manufacturing industries, energy and water supply, sanitation activities, waste management and decontamination),
    \item[(b)]  \textbf{construction},
    \item[(c)] \textbf{commerce} (including wholesale, retail trade, repair of motor vehicles and motorcycles, transport and storage, hostelry) and
    \item[(d)]\textbf{services} (including communication, financial and insurance services, administrative activities and support services, education,  health and social services, artistic, recreational and entertainment activities) 
\end{enumerate} 
according to its main economic activity with reference date 1st January 2022.

\begin{figure}[h!]
    \centering
      \includegraphics[scale=.89]{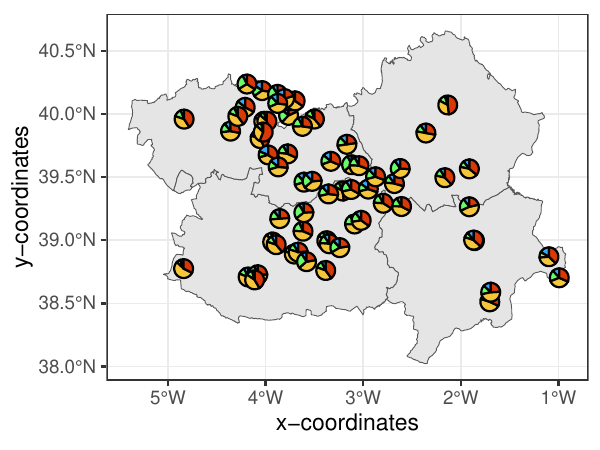}
   \caption{Distribution of 4-part business sector composition on the Spanish Plateau for municipalities  with at least 1000  inhabitants. Pie charts show the local decomposition of the economy into the business sector proportions of industry (blue), construction (green), commerce (orange) and service (red).}
\label{fig:maplabour}
\end{figure}

Next, restricting the data to the Spanish provinces of \textit{Albacete}, \textit{Cuenca}, \textit{Cuidad Real} and \textit{Toledo} yielded a sample of 66 municipalities with complete information on the business sector  decomposition. The four selected provinces are located southeast of Madrid and belong to the geographic  region of  \textit{La Mancha}, the Spanish Plateau which is characterised by a homogeneous climate and strong similarity in its local population densities. Due to these characteristics, \textit{La Mancha} has been used as a particular example of a homogeneous spatial point process in the literature  \citep[see e.g.][]{Glass1971, ripley77,Chiu2013}. All collected information was then considered as a marked  spatial point process by treating the attached coordinates as points and the closed four-part composition as point attribute. A visualisation of the point pattern at hand with the corresponding composition-valued marks shown as  pie charts is depicted in Figure \ref{fig:maplabour}.
The observed configuration of the marked points reflects a clear tendency of clustering for the point locations in combination with some variation over the individual pie charts, which indicate a clear predominance of the \textit{commerce} and \textit{service} sectors within the four-part compositions. While the heterogeneity of the business sector  decomposition seems to be maximal at larger interpoint distances, the composition seems to become more homogeneous for closely neighbouring points. 
\begin{table}[h!]
    \centering
        \begin{tabular}{lrrrrrr}
\hline
    Sector (in \%) & Min & 1st quantile & Mean & Median & 3rd quantile & Max\\
        \hline
Industry &  2.81 &  7.35 & 8.83 & 9.56 & 11.72 & 19.78 \\ 
Construction  & 6.69  & 11.96  & 15.61 & 15.90 & 18.49 & 29.30\\          
Commerce & 31.15 & 38.28 & 41.73 & 42.02 & 44.96 & 56.07\\
Service & 18.18 & 27.11 & 31.47 & 32.52 & 37.08 & 56.72\\
  \hline
  \end{tabular}
    \caption{Summary statistics for the closed 4-part business sector  composition computed from the Spanish business sector  data for 66 municipalities with at least 1000 inhabitants on the Spanish Plateau.}
    \label{tab:summaryab}
\end{table}

This observed variation of the marks is also supported by the numerical summary statistics of the business sector  proportions reported in Table \ref{tab:summaryab}, which again reflect a clear predominance of the commerce and service sectors contrasted with only smaller proportions of the industry and construction sectors. The geometric means of the four parts highlight clear differences between the sectors \textit{industry}  (0.09), \textit{construction} (0.16),  \textit{commerce} (0.43) and \textit{services} (0.32). 

\begin{figure}[t!]
    \centering
    \includegraphics[scale=0.65]{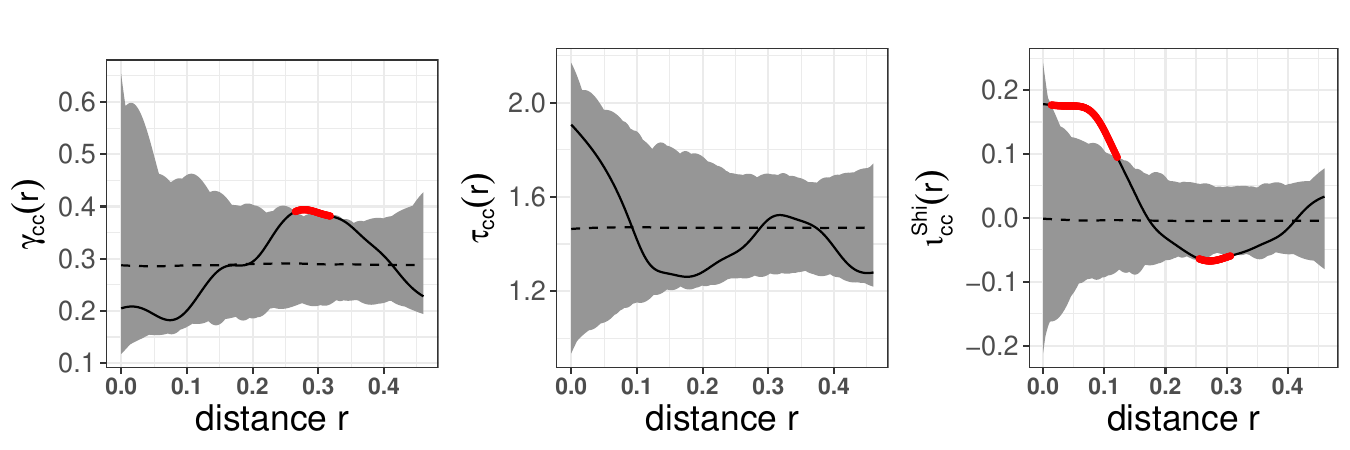}
    \caption{Compositional mark summary characteristics computed from the 4-part economic sector compositions: Mark variogram $\gamma_{\cb\cb}$ (left), the conditional mean of the product of marks $\tau_{\cb\cb}$ (central) and Shimantani's $\iota_{\cb\cb}^{\mathrm{Shi}}$ (right).
    The grey areas are 95\% global envelopes constructed from 3000 simulations under the random labeling hypothesis. The dashed line corresponds to the mean function under the random labeling hypothesis and the solid curve to the test function on the observed data.
    }
    \label{fig:mark.res:lab4eco}
\end{figure}

For the composition-valued marks, we  computed the same  three compositional mark summary characteristics with global envelopes as before to investigate the spatial joint variation, association and autocorrelation of the complete 4-part composition (see Figure \ref{fig:mark.res:lab4eco}). While global envelopes for the mark variogram $\gamma_{\cb\cb}$ (left) and Shimantani's $\iota_{\cb\cb}^{\mathrm{Shi}}$ (right) suggest  deviations from the independent mark assumption for some distances $r$, the conditional mean of the inner  product of marks $\tau_{\cb\cb}$ (central) is completely covered by the global envelopes. 
The mark variogram $\gamma_{\cb\cb}$ indicates that the average dispersion between the transformed 4-part compositions of any pairs of points is greater than expected under the independent mark assumption at distances of around 0.3 units. 
For distances $r\leq 0.2$ units, the mark variogram is smaller than expected under the random labeling hypothesis, although the empirical function stays within the envelope. This suggests similarity of the marks for pairs of nearby municipalities 
(although not significant), 
with increasing  variability as the distances between point locations become larger, at least until about 0.3 units, where there is the most data. This might be explained by a strong variation in and clustering of the contribution of sectors such as e.g.\ tourism or banking to the local economy which, in turn, would affect the relative size of the service and commerce sectors.
For Shimantani's $\iota_{\cb\cb}^{\mathrm{Shi}}$ the findings suggest positive conditional  spatial  autocorrelation of the compositions at any nearby points with $r< 0.15$ units, and negative autocorrelation for $r\approx 0.3$ units, consistent with the results of the variogram. 

\begin{figure}[t!]
    \centering
      \includegraphics[scale=.65]{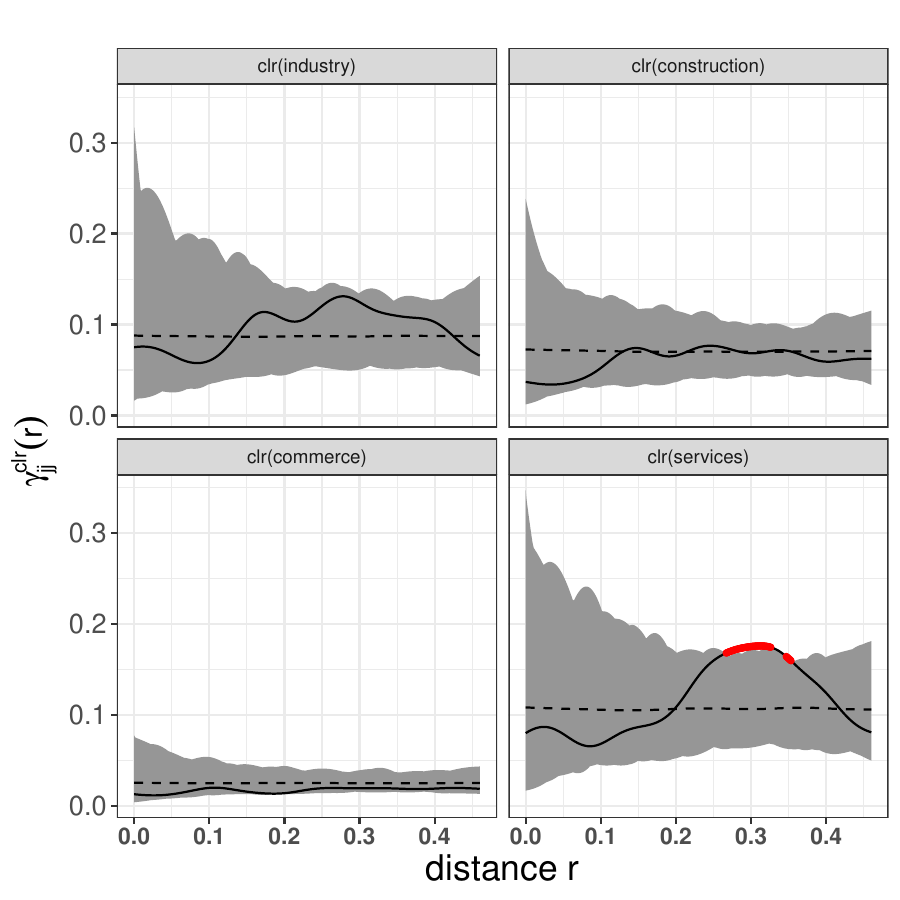}
   \caption{Componentwise mark variograms $\gamma_{jj}^{\mathrm{clr}}$ computed from the clr transformed 4-part economic sector composition. 
   The grey bands represent 95\% global envelopes constructed from 3000 simulations under the random labeling hypothesis. The dashed line corresponds to the mean function under the random labeling hypothesis and the solid curve to the test function on the observed data. 
   }
\label{fig:results:tree:total:pw}
\end{figure}

\begin{figure}[t!]
    \centering
      \includegraphics[scale=.65]{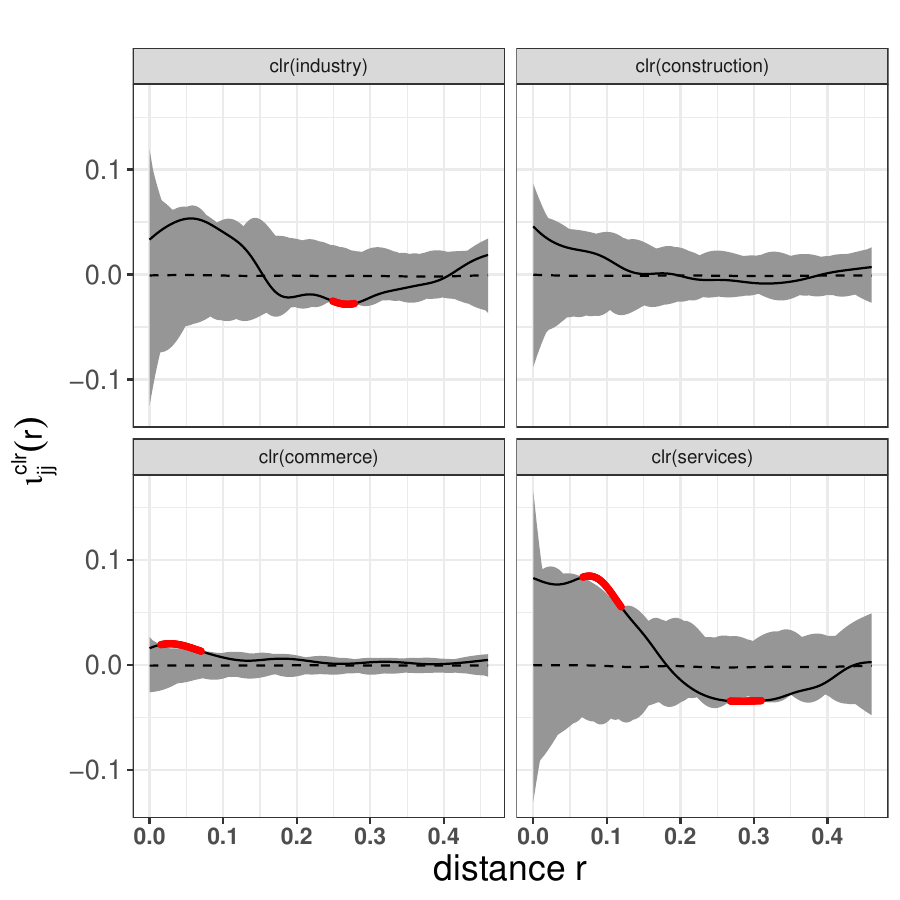}
   \caption{Componentwise Shimantani's $\iota_{jj}^{\mathrm{clr}, \mathrm{Shi}}$ and envelopes based on 3000 simulations computed from the clr transformed  4-part economic sector compositions.  The dashed line corresponds to the mean function under the random labeling hypothesis and the solid curve to the test function on the observed data.}
\label{fig:results:tree:total:iota:pw}
\end{figure}

To investigate the effect of each of the four parts on the above results, we additionally computed the componentwise clr mark variograms  $\gamma^{\mathrm{clr}}_{jj}$ and componentwise Shimantani's $\iota_{jj}^{\mathrm{clr}, \mathrm{Shi}}$ functions and plotted the results. Of all four $\gamma^{\clr}_{jj}$ functions depicted in Figure \ref{fig:results:tree:total:pw}, only the mark variogram of clr(services) highlights deviations from the independent mark setting. This would suggest that the service sector proportions are more heterogeneous at larger distances which is consistent with the visual  impression from Figure \ref{fig:maplabour}.  
By contrast, except for the clr-transformed construction sector, parts of all empirical $\iota_{jj}^{\clr, \mathrm{Shi}}$ functions are outside the $95\%$ envelopes for some distances $r$ (see Figure \ref{fig:results:tree:total:iota:pw}). While we found positive autocorrelations for clr(commerce) and clr(services) at smaller distances $r \approx 0.1$ units, both clr(industry) and clr(services) are below the $95\%$ envelopes at distances $r \approx 0.3$ units corresponding to negative autocorrelation. This again would relate to similarity among the business sectors for closely neighbouring municipalities and an increasing heterogeneity with increasing distances, with some differences in (the strength of) this pattern between sectors.

\section{Conclusion}\label{sec:conclusio}

Combining methodological concepts for compositional data and spatial point processes, this paper introduces a novel class of composition-valued marked spatial point processes. Our proposed set 
of different (functional) mark summary characteristics 
allows to decide on the mark independence assumption and investigate the pairwise dependencies of a new type of marked spatial point process. All developments are formalized through extended test functions, which generalise well-known interpretations to the present context. Transforming the composition-valued marks to the Euclidean space, the proposed tools can build on established methods for real-valued marks and can borrow strength from existing computational implementations. 

Allowing for the characterisation of the spatial variation and association between both the complete composition-valued marks as well as their distinct compositional parts, the proposed  extensions are helpful tools to highlight different aspects of the mark pattern. While the overall measures characterise the global interdependencies of the marks as a function of the interpoint distance, their componentwise counterparts provide useful insights into the contribution of the distinct parts to the overall results. 

Apart from methods for purely composition-valued marks, we covered extensions to mixed marks including both composition-valued and absolute information. 
These methods also allow to decompose vector-valued marks into the absolute and the relative information contained therein, highlighting patterns also in the relative information and avoiding that results for the vector-valued marks are mainly driven by the absolute information. For a combined analysis of both pieces of information, we introduced weights into the overall scalar product, which control for the differences in variation  between both types of marks \citep{HappGrevenJASA}. We have here covered the case of spatial point processes with composition-valued marks, which can be seen as a special case of spatial point processes with more general object-valued marks.

Further extensions in this direction might also include alternative non-scalar marks such as  density-valued or shape-valued marks. Note that as a by-product of our developed methods, we also showed how to handle vector-valued marks and derived both componentwise and (full-vector) compositional summary characteristics as well as their relationship, a result which is of independent interest in its own right. 

\section*{Acknowledgements}

The authors gratefully acknowledge financial support through the German Research Foundation and Research Council of Finland. Matthias Eckardt and Sonja Greven were funded by the Deutsche Forschungsgemeinschaft (DFG, German Research Foundation) - project numbers  467634837 (Walter Benjamin grant for ME) and 459422098 (SG), respectively. 
Mari Myllym{\"a}ki was financially supported by the Research Council of Finland (Grant numbers 295100, 327211) and
the European Union -- NextGenerationEU in the Research Council of Finland project (Grant number 348154) under flagship ecosystem for Forest-Human-Machine Interplay -- Building Resilience, Redefining Value Networks and Enabling Meaningful Experiences (UNITE) (Grant numbers 337655 and 357909).


\newpage

\section*{Supplement of On spatial point processes with composition-valued marks}

\subsection*{Transformations in the presence of structural zeros}\label{sec:zeros}
 Where zeros are not structural (genuine) but e.g.\ due to small samples or rounding, 
they are commonly replaced or imputed from the data \citep[see e.g.][]{MartinFernandez2003, LUBBE2021104248}. By contrast, general  approaches for structural zeros  include amalgamations of the zero components into larger nonzero groups, separate treatment of zero and  nonzero components, or restriction of the data to completely observed compositions only. Alternative approaches for structural zeros include representations of the data in linear or nonlinear geometric spaces through either combinations of log- and power transformations or projections onto the sphere using square root transformations. In the linear geometry framework, early contributions include the folded power  \citep{Atkinson:1985} and Box-Cox \citep{10.5555/17272, https://doi.org/10.1002/cem.1180050310} transformations, which both tend to the $\alr$-transformation if the power parameter $p\rightarrow 0$. However, these transformations are problematic for some samples and certain properties of the composition are not well preserved \citep{Barcelo1996}.  Instead, the $\alpha$-transformation introduced by  \cite{alpha:2011} and further investigated by 
\citet{tsagris2015} and \citet{Tsagris2016} is defined through the map $\alpha:\mathds{S}^D\rightarrow \mathds{R}^{D-1}$    with
$\alpha(\cb)= \h\mathbf{u}_{\alpha}$ where $\mathbf{u}_{\alpha}=(D\cdot(\cls(\alpha \odot\cb)-\mathds{1}_D))/\alpha$. The $\alpha$-transformation tends to $\ilr$-coordinates when $\alpha\rightarrow 0$ and a linear transformation if  $\alpha\rightarrow 1$.  We note that similar ideas were also developed by   \cite{Greenacre2009,GREENACRE20093107}  within the context of correspondence analysis. While the $\alpha$-transformation is well-defined for $\alpha>0$, it maps the data into a codomain that is  a subspace $\mathds{A}_{\alpha}^{D-1}$ of $\mathds{R}^{D-1}$ with $\lim_{\alpha\rightarrow 0}\mathds{A}_{\alpha}^{D-1}\rightarrow \mathds{R}^{D-1}$ given by  
\[
\mathds{A}^{D-1}=\left\lbrace\h\mathbf{u}_{\alpha} \bigg | -\frac{1}{\alpha} \leq u_{j,\alpha} \leq \frac{D-1}{\alpha},\sum_{j=1}^Du_{j, \alpha}=0 \right\rbrace
\]
 \citep{https://doi.org/10.1111/anzs.12289}. To overcome this limitation, \cite{CLAROTTO2022100570} proposed to use a centered $\alpha$ -transformation  $(\aclr)$,
\[
\aclr(\cb)=\left(\alpha^{-1}\left(c_1^{\alpha}-\frac{1}{D}\sum_{j=1}^Dc_j^{\alpha}\right),\cdots,\alpha^{-1}\left(c_D^{\alpha}-\frac{1}{D}\sum_{j=1}^Dc_j^{\alpha}\right)\right)
\]
with sum-to-zero constraint
or an isometric $\alpha$-transformation ($\ailr$) where $\ailr(\cb)=
\aclr(\cb)\h^{\top}$ and 
$\aclr(\cb)=\alpha^{-1}(\mathbf{G}_\mathrm{D}\cb^\alpha)$. Both, 
$\ailr$ and $\aclr$ tend to $\ilr$ and $\clr$, respectively,  when $\alpha\rightarrow 0$.

Different from the linear space formulation, some parts of the literature considered square root transformations, which project the data onto a $(D-1)$-dimensional (hyper-)sphere. While any such transformation allows for structural zeros,  it imposes a nonlinear geometry and knowledge in directional data analysis  techniques is required to interpret the results \citep{https://doi.org/10.1111/j.1467-9868.2010.00766.x, doi:10.1080/01621459.2014.990563,WANG2007459}.

\subsection*{Extensions to multitype point  processes with  two distinct composition-valued marks}\label{sec:CodaMulti}

While the above methods considered the analysis of univariate point processes with one composition-valued mark, we next discuss extensions to $k$-variate point processes $(\mathbf{x}_1,\ldots, \mathbf{x}_k)$ with two distinct associated composition-valued marks $(\cb^a,\cb^b)$ on $\R^2\times\si^{D_1}\times\si^{D_2}$ such that each  component $\mathbf{x}_k=\{x_i, (\cb^a(x_i),\cb^b(x_i))\}_{i=1}^{n_k}$ is a set of $n_k$ points with two compositional marks. 
Such extended characteristics might be useful tools to investigate e.g. the association of one composition-valued mark, say a base-to-trunk composition, for different tree species and also to explore the associations and distributional characteristics of different compositions over space, say a tree and a soil composition. Extending the above methods to this setting not only allows to highlight particular aspects of the spatial behaviour  of one composition, but also the cross-association or cross-variation between different components of a multitype point process and/or different compositions.

Extending \eqref{eq:componentwise:meanprodmarks:generic} to two distinct compositions yields a generic cross-composition conditional expectation of the product of $\psi$-transformed marks  $\tau^{\psi, ab}_{jl}$,
\begin{equation}\label{eq:cross:comp:tau}
 \tau^{\psi, ab}_{jl}(r)=\mathds{E}_{\circ,r}\left[\psi_j(\cb^a(\circ))\psi_l(\cb^b(\mathbf{r}))\right]   
\end{equation}
where $\psi_{j}(\cb^a(\circ))$ and $\psi_l(\cb^b(\mathbf{r}))$ are the $j$-th and $l$-th elements of the $\psi$-transformed compositions $\cb^a$ and $\cb^b$. Under independence of the two compositions at distance $r$, \eqref{eq:cross:comp:tau} tends to the product of means $\mu_j^{\psi,a}\cdot \mu_l^{\psi,b}$. Similarly, reformulation of \eqref{eq:componentwise:vario:generic} leads to a generic cross-composition mark variogram $\gamma^{\psi, ab}_{jl}(r)$ as  
\begin{equation}\label{eq:cross:comp:vario}
\gamma^{\psi, ab}_{jl}(r)=\mathds{E}_{\circ,r}\left[0.5\cdot\left(\psi_j(\cb^a(\circ))-\psi_l(\cb^b(\mathbf{r}))\right)^2\right].
\end{equation}

In settings where there are additional integer-valued marks, e.g. the kind of tree (spruce, beech etc.), i.e.\ the process can be considered as multivariate, we can extend   
\eqref{eq:cross:comp:tau} above to integer-valued marks, say points of type $p$ and $q$, yielding a cross-composition cross-type mean product of marks $\tau^{ab, pq}_{jl,hw}$, 
\begin{equation}\label{eq:cross:comp:type:tau}
 \tau^{\psi,ab}_{jl,pq}(r)=\mathds{E}_{\circ,r}\left[\psi_j(\cb^a_{p}(\circ))\psi_l(\cb^b_{q}(\mathbf{r})\right],  
\end{equation}
where $\psi_j(\cb^a_{p})$ and $\psi_j(\cb^b_{q})$ are the $j$-th part, respectively $l$-th part, of the $\psi$-transformed compositions $\cb^a$ and $\cb^b$ for points of type $p$ and $q$, respectively.

\newpage
\bibliographystyle{ecta}
\bibliography{fmspp}

\end{document}